\documentclass[12pt,a4paper]{article}
\usepackage{latexsym,graphicx,multirow}
\usepackage{float}
\usepackage{amssymb}
\usepackage{amscd}
\usepackage{amsthm}
\usepackage[left=2cm,top=2cm,right=1.8cm,bottom=2cm]{geometry}
\usepackage[hyperfootnotes=false]{hyperref}
\usepackage{epstopdf}
\usepackage{cite}
\usepackage{url}
\usepackage{orcidlink}
\usepackage[utf8]{inputenc}
\usepackage{enumerate}
\usepackage{caption}
\usepackage{subcaption}
\usepackage{amsmath}
\usepackage[mathscr]{euscript}
\usepackage{calrsfs}
\usepackage{framed}
\usepackage{xparse}
\usepackage{fancyvrb}
\usepackage{lscape}
\usepackage{tikz}
\usetikzlibrary{patterns.meta}
\usepackage{multirow}
\usepackage{latexsym,graphicx,multirow}
\theoremstyle{plain}
\usepackage{longtable}
\usetikzlibrary{positioning,shapes}
\usepackage{algorithm}
\usepackage{rotating}
\usepackage{algpseudocode}
\newtheorem{theorem}{Theorem}[section]
\newtheorem{corollary}{Corrolary}[section]
\newtheorem{proposition}{Proposition}[section]
\newtheorem{remark}{Remark}[section]
\newtheorem{lemma}{Lemma}[section]
\newtheorem{definition}{Definition}[section]

\newcommand{\be}{\begin{equation}}
	\newcommand{\ee}{\end{equation}}
\newcommand{\bp}{\begin{proposition}}
	\newcommand{\ep}{\end{proposition}}
\newcommand{\ben}{\begin{equation*}}
	\newcommand{\een}{\end{equation*}}
\newcommand{\bd}{\begin{definition}}
	\newcommand{\ed}{\end{definition}}
\newcommand{\bl}{\begin{lemma}}
	\newcommand{\el}{\end{lemma}}
\newcommand{\bn}{\begin{notation}}
	\newcommand{\en}{\end{notation}}
\newcommand{\bcon}{\begin{construction}}
	\newcommand{\econ}{\end{construction}}
\newcommand{\bea}{\begin{eqnarray}}
	\newcommand{\eea}{\end{eqnarray}}
\newcommand{\bee}{\begin{eqnarray*}}
	\newcommand{\eee}{\end{eqnarray*}}
\newcommand{\bt}{\begin{theorem}}
	\newcommand{\et}{\end{theorem}}
\newcommand{\br}{\begin{remark}}
	\newcommand{\er}{\end{remark}}
\newcommand{\bo}{\begin{observation}}
	\newcommand{\eo}{\end{observation}}
\newcommand{\bex}{\begin{example}}
	\newcommand{\eex}{\end{example}}
\newcommand{\bc}{\begin{corollary}}
	\newcommand{\ec}{\end{corollary}}
\newcommand{\numsize}[1]{{\fontsize{11pt}{13pt}\selectfont #1}}
\usepackage{array}
\newcolumntype{N}{>{\numsize}c}
\NewDocumentCommand{\INTERVALINNARDS}{ m m }{
	#1 {,} #2
}

\makeatother
\NewDocumentCommand{\interval}{ s m >{\SplitArgument{1}{,}}m m o }{
	\IfBooleanTF{#1}{
		\left#2 \INTERVALINNARDS #3 \right#4
	}{
		\IfValueTF{#5}{
			#5{#2} \INTERVALINNARDS #3 #5{#4}
		}{
			#2 \INTERVALINNARDS #3 #4
		}
	}
}
\usepackage[symbol]{footmisc}
\renewcommand{\thefootnote}{\fnsymbol{footnote}}
\begin{document}	
	\begin{center}
		\large{\bf{Interval Estimation of the Common Shape Parameter and Coefficient of Variation of Several Weibull Populations under Progressive Censoring}} \\
		\vspace{5mm}
		\normalsize{ Bankitdor M. Nongrum$^{1}$\orcidlink{0009-0003-8828-5046}, Adarsha Kumar Jena$^{2}$\footnote[1]{Corresponding author.} \let\thefootnote\relax\footnote{\textit{E-mail address:} jadarsha@gmail.com (Adarsha Kumar Jena)} \orcidlink{0000-0001-8372-8176}
		} \\
		\vspace{5mm}
		\normalsize{$^{1,2}$Department of Mathematics, National Institute of Technology Meghalaya, Sohra (Cherrapunji) - 793108, Meghalaya, India}\\
	\end{center}
	\begin{abstract}
	The Weibull distribution is one of the most flexible continuous probability distributions used to model various failure rates and skewed data in reliability engineering, industry, weather studies and cancer studies. It is a common scenario in statistical inference that several Weibull populations share the same shape parameter, which also implies that they have the same coefficient of variation. While the inferential study of the common shape parameter is often considered for complete samples, the presence of censored data requires a separate investigation that has not received enough attention in the existing literature. Therefore, the focus of this article is on a comparative study of interval estimators for the common shape parameter and the common coefficient of variation under progressive type-II censoring using frequentist methods based on large-sample theory, variance estimates recovery, generalized pivots, and Bayesian inference. An optimal censoring scheme is also proposed to enhance the robustness of interval estimation. Numerical data analyses using a simulation study and a real carbon fiber strength data example are carried out for comparison, and the results recommend the intervals based on Bayesian and variance estimates recovery methods for their satisfactory performance.
	\end{abstract}
	\smallskip 
	{\bf Keywords}: Common Shape Parameter, Weibull distribution, Progressive Type-II Censoring, Confidence Intervals, Optimal Censoring
	
	\smallskip 
	\noindent{\bf \textit{2020 Mathematics Subject Classification}}: 62F10,	62F15, 62N02, 62N05 
	\section{Introduction}\label{sec1}

In many practical studies, the time-to-failure or time-to-event does not always follow a normal distribution. Conventional yet simple models such as the exponential distribution, which assumes a constant failure rate, are unrealistic for most physical systems that exhibit decreasing or increasing failure rates. Practitioners in many interdisciplinary fields require a more flexible mathematical model that can represent asymmetric, non-normal data and handle early-, late-stage, or random failures within a single functional framework. The Weibull distribution is one such flexible continuous probability distribution that is used to model various types of failure mechanisms and skewed data in reliability engineering and survival analysis. It has a variety of applications, ranging from management \cite{DP} to weather studies \cite{MA}, industries \cite{JY}, reliability theory \cite{SF}, survival analysis and cancer studies \cite{AH,MN}. \\
\indent In practice, the cost, time, risk, missing information, and invalid observations often constrain the collection of complete data. As a result, the data frequently obtained in many applications such as survival analysis and reliability theory are subject to censoring. Among the various censoring mechanisms, the progressive type-II censoring scheme is one of the most widely applied. Suppose that $n$ experimental units are placed on test initially at $x_0 = 0$. A predetermined number $m$ of failure events are completely observed, with corresponding ordered failures recorded at time points $x_1, x_2, \ldots, x_m$. At the occurrence of the $i^\text{th}$ failure, $i = 1, 2, \ldots, m$, some $R_i$ surviving units are randomly removed from the experiment. Let $\mathbf{R} = (R_1, R_2, \ldots, R_m)$. When $R_i = 0~\forall~i = 1,2,\ldots,m-1$, $R_m\neq0$, then the scheme reduces to the type-II censoring scheme. If furthermore, $R_m=0$, then none of the surviving units are removed during the experiment. This implies that all units are observed until failure, which reduces to the complete sampling case ($m=n$). Thus, the progressive type-II censoring scheme allows units to be removed at different stages of the experiment without removing the necessary information. Therefore, it is particularly useful in reliability studies because it provides greater flexibility than conventional censoring schemes such as type-II or right censoring. For a comprehensive study of the progressive type-II censoring scheme and its statistical properties, refer to \cite{NB}. \\
\indent The inference on common parameters or characteristics of several populations has gained traction in recent years due to practical relevance, as this enables information from multiple populations to be combined. Existing studies on Weibull populations have all considered a complete sampling scenario, and primarily focused on estimation of common shape parameters or mean \cite{MLO}. Under the progressive type-II censoring scheme, the only related work that exists in the literature is that of \cite{HK}, which deals with the estimation of quantiles with a common location parameter for several exponential distributions. In many applications, the assumption of equality of the shape parameter for several Weibull populations is highly significant. For example, the testing of batches of components in industry produced under similar manufacturing treatment is often considered, such as in rolling contact endurance tests \cite{JIM}. It is fairly common to assume that even if their characteristic lifetimes differ, they usually share the same failure mechanism represented by the common shape parameter \cite{WN,JFL}. The maximum likelihood estimation approach is frequently used to estimate the common shape parameter of several Weibull distributions \cite{JIM2,EH}, despite lacking a closed-form formulation and producing highly biased estimates. Since an unbiased estimator with explicit form is unattainable, researchers have resorted to bias-correction of the maximum likelihood estimator \cite{ZY,YS}. For the case of two Weibull populations with the same shape parameter, several inferential works have been carried out, see \cite{NA,HZ,SM,FS}. \\
\indent The coefficient of variation is a useful statistical tool for comparing the relative variability of several datasets or populations. It is mathematically defined as the ratio of the standard deviation relative to the mean. The coefficient of variation is scale-free and unit-free, and is often preferred over the standard deviation or variance for comparing relative dispersion of datasets with different scales, means, or units. It has been widely used as an index of measurement reliability \cite{OS}. It has vast applications ranging from weather studies \cite{QY} to medical studies \cite{GJG} and reliability theory \cite{OOB}. For the Weibull distribution, the coefficient of variation is a function of the shape parameter itself. Therefore, it draws the conclusion that if several populations are governed by the same shape parameter, they also share the same coefficient of variation. In the existing literature, the statistical inference for a common coefficient of variation for some well-known distributions has gathered significant attention for its applications and relevance. Following the homogeneity hypothesis \cite{SEA}, \cite{LT} pioneered the inferential study of a common coefficient of variation using generalized pivotal quantities. \cite{SV} provided the confidence intervals of a common coefficient of variation and hypothesis testing for several normal populations. Similar works for the common coefficient of variation for several normal distributions have been carried out by \cite{MRK,WTNorm}. For the shifted exponential distribution, some confidence interval estimation methods based on large samples, variance estimates recovery, and generalized confidence intervals have been proposed for the weighted coefficient of variation  \cite{WTExp}. Similarly the inferential study for the common coefficient of variation of several lognormal distributions, has been considered in \cite{WTSNR,WTCV}; for several gamma distributions in \cite{PS}; and for several Birnbaum-Saunders distributions see \cite{UJ}.\\
\indent From this comprehensive review of the literature, it is observed that the problem of estimating a common shape parameter for several Weibull populations under any form of censoring has not been investigated yet. This is a highly significant methodological gap, as the shape parameter is of fundamental interest in reliability analysis. The statistical inference for the Weibull distribution is considerably more challenging under censoring than in the complete-sample setting. Not only does the former result in a substantially more complicated likelihood structure, but the existing methodologies are not applicable without significant modification. Furthermore, it is a statistical fact that interval estimators are known to provide better precision than point estimators due to their accounting of sampling errors. Therefore, in this study, the interval inference for the common shape parameter and coefficient of variation of several Weibull populations based on progressive type-II censored samples is considered. Since this type of censoring scheme generalizes the complete sampling plan, this work and its further extensions generalizes some of the results by \cite{MLo1,MLB}.  Therefore, the proposed methodology extends the existing body of work on single and common-parameter estimation to a setting that has not previously been explored. Although there is a rich literature on progressive censoring, this study addresses a theoretical problem of independent statistical interest. The results of this study therefore contribute both to the theory of common-parameter estimation under censoring and to its application in modern reliability and life-testing experiments. \\
\indent The remainder of this article is outlined as follows. In Section 2, the interval estimation methods are discussed. Each method is organized in different subsections, which include the methods of large samples, variance estimates recovery, generalized pivots, and Bayesian methods. In Section 3, an optimal censoring design is proposed for enhancing robust estimation. A numerical Monte Carlo simulation study is conducted and discussed in Section 4. A real data set analysis is also considered for illustrative purposes in Section 5.
\section{Interval Estimation Methods}\label{sec2}
Consider the $k$ progressive type-II censored samples denoted as $\{(X_{1j:m_j:n_j}, R_{1j}),(X_{2j:m_j:n_j},R_{2j}),...,$ $(X_{m_jj:m_j:n_j}, R_{m_jj})\}_{j=1}^{k}$ from several Weibull populations with respective rate parameters $\lambda_j>0$, and a common shape parameter $\eta > 0$. The probability density function for the random variable $X_j$ from the $j^\text{th}$ population is defined by:
\begin{equation} \label{2.1}\tag{2.1}
	f(x_{j}|\eta,\lambda_j) = \begin{cases}
		\eta \lambda_j x_{j}^{\eta-1}\exp\left(-\lambda_jx_j^\eta\right),\quad &x_j > 0 \\
		0, \quad &\text{otherwise},
	\end{cases}
\end{equation}
while the reliability function is given by
\begin{equation} \label{2.2}\tag{2.2}
	S(x_j|\eta,\lambda_j) = \begin{cases}
		\exp\left(-\lambda_jx_j^\eta\right),\quad &x_j > 0 \\
		0, \quad &\text{otherwise}.
	\end{cases}
\end{equation}
This parametrization allows for computational and algebraic simplicity. The mean ($\mathbb{E}\left[X_j\right]$) and variance ($\text{Var}\left[X_j\right]$) of $X_j$ are respectively given as follows:
\begin{equation} \nonumber
	\mathbb{E}\left[X_j\right] = \frac{1}{\lambda_j^{1/\eta}} \Gamma\left(1+\frac{1}{\eta}\right),\quad\&\quad \text{Var}\left[X_j\right] = \frac{1}{\lambda_j^{2/\eta}}\left[\Gamma\left(1+\frac{2}{\eta}\right)-\Gamma^2\left(1+\frac{1}{\eta}\right)\right].
\end{equation}
Therefore, the coefficient of variation is calculated as follows:
\begin{equation} \label{2.3}\tag{2.3}
	\text{C}_\eta = \frac{\sqrt{\text{Var}\left[X_j\right]}}{\mathbb{E}\left[X_j\right]} = \sqrt{\frac{\Gamma\left(1+\frac{2}{\eta}\right)}{\Gamma^2\left(1+\frac{1}{\eta}\right)} - 1}.
\end{equation}
which shows $\text{C}_\eta$ is a function of $\eta$ alone and is independent of $j$. Hence, if several Weibull distributions share the same shape parameter, then they also have a common coefficient of variation. Now, rather than computing the joint likelihood function of $k$ samples, the individual likelihood for the $j^{\text{th}}$ sample from the population Weibull$\left(\eta,\lambda_j\right)$ is instead defined:
\begin{equation} \label{2.4}\tag{2.4}
	\mathscr{L}\left(\eta,\lambda_j\right) = C_j\prod_{i=1}^{m_j} f\left(x_{ij}\right) \left[S\left(x_{ij}\right)\right]^{R_{ij}},
\end{equation}
where, $C_j= \prod\limits_{i=1}^{m_j-1}\left[n_j - \sum_{l=1}^{i}\left(R_{lj}+1\right)\right]$. This approach, which is not only computationally simpler, but is also required for the proposed methodologies in this article. The log-likelihood function of the $j^{\text{th}}$ sample is given by 
\begin{equation} \label{2.5}\tag{2.5}
	\ell_j = m_j \log \eta + m_j \log \lambda_j + \left(\eta - 1\right)\sum_{i=1}^{m_j}\log x_{ij:m_j:n_j} - \lambda_j\sum_{i=1}^{m_j} \left(R_{ij}+1\right)x_{ij:m_j:n_j}^\eta.
\end{equation}
The maximum likelihood estimator (MLE) of the shape parameter $\eta$ using the $j^\text{th}$ progressive type-II censored sample is denoted by $\hat{\eta}_j$, and is obtained by solving the following equation using root-solving techniques:
\begin{equation}
	\frac{m_j}{\hat{\eta}_j} + \sum_{i=1}^{m_j}\log x_{ij:m_j:n_j} -  \frac{m_j\sum_{i=1}^{m_j} \left(R_{ij} + 1\right)x_{ij:m_j:n_j}^{\hat{\eta}_j}\log x_{ij:m_j:n_j}}{\sum_{i=1}^{m_j} \left(R_{ij} + 1\right)x_{ij:m_j:n_j}^{\hat{\eta}_j}} = 0,\quad j=1,2,...,k. \label{2.6}\tag{2.6}
\end{equation}
By the property of invariance, the MLE of $\text{C}_\eta$ is $\hat{\text{C}}_\eta$. \cite{MT} have computed the expected Fisher's information matrix by considering the scale parameter as $\lambda_j^{-1/\eta}$. However, in this paper the reparametrization of the scale to the rate parameter allows for restatement of their theorem to obtain the Fisher's information matrix as follows:
\begin{theorem} \label{Th-2.1}
	For the $j^\text{th}$ progressive type-II censored sample $\{\left(X_{1j:m_j:n_j}, R_{1j}\right),\left(X_{2j:m_j:n_j},R_{2j}\right),$ $...,\left(X_{m_jj:m_j:n_j}, R_{m_jj}\right)\}_{j=1}^{k}$, from the Weibull distribution with joint density function $f\left(x_{j}|\eta,\lambda_j\right)$, the entries of the expected Fisher's information matrix $\mathscr{I}_j$ are given by:
	\begin{align}
		\mathscr{I}_j^{11} &= \frac{1}{\eta^2} \sum_{r_j=0}^{m_j-1}C_{r_j}\sum_{l=0}^{r_j} \frac{D_l^{(r_j)}}{R_l^{(r_j)}}\left[\left\{1-\left(\gamma + \log \lambda_j + \log R_l^{(r_j)}\right)\right\}^2 + \frac{\pi^2}{6}\right], \nonumber \\
		\mathscr{I}_j^{12} &=	\mathscr{I}_j^{21} = \frac{1}{\eta\lambda_j}\sum_{r_j=0}^{m_j-1}C_{r_j}\sum_{l=0}^{r_j} \frac{D_l^{(r_j)}}{R_l^{(r_j)}}\left[1-\left(\gamma + \log \lambda_j + \log R_l^{(r_j)}\right)\right], \nonumber \\
		\mathscr{I}_j^{22} &= \frac{m_j}{\eta^2}, \nonumber
	\end{align}
	where, $D_l^{(r_j)}=\left(-1\right)^l/\left[\prod_{s=1}^{l}\sum_{u=r_j-l+1}^{r_j-l+s}\left(R_{uj}+1\right)\right]\left[\prod_{s=1}^{r_j-l}\sum_{u=s}^{r_j-l}\left(R_{uj}+1\right)\right]$, $R_l^{(r_j)} = n_j - r_j + l - \sum_{s=1}^{r_j - l} R_{sj}$, and $\gamma$ is the Euler-Mascheroni constant.
	\begin{proof}
		\normalfont The proof is similar to Theorem 3 of \cite{MT}, where instead the following type of integral is used:
		\begin{equation}
			I(q,\omega,R,\eta,\lambda_j) =\int_{0}^{\infty}[x^\eta]^q [\log x]^\omega f(x;\eta,\lambda_j) S(x)^{R-1}dx, \tag{2.7}\label{2.7}
		\end{equation}
		where, $q=0,1$ and $\omega=0,1,2,3$.
	\end{proof}
\end{theorem}
The inverse of $\mathscr{I}_j$ computed at the given MLEs results in the asymptotic covariance matrix denoted by $\widehat{V}_j$. However, the quantity $C_{r_j} = O(n_j^{r_j+1})$ can explode with larger $n_j$, therefore accumulating floating-point errors that result in large, inaccurate entries of Fisher's information matrix. Sometimes, it becomes essential to drop the expectation altogether and compute the observed Fisher's information matrix directly. In the following subsections, the various methods of interval estimation are discussed.
\subsection{Large-Sample Method}
Large sample intervals are based on the asymptotic normality property of estimators. For example, the MLE exhibits this property and is thereby significantly utilized in statistical inference for point and interval estimation. According to \cite{FAG}, the large sample estimate of the function $\phi(\eta) = \eta$ and $\text{C}_\eta$, is a pooled-in estimator as given below:
\begin{equation} \label{2.8} \tag{2.8}
	\phi(\hat{\eta}) = \sum_{j=1}^{k} \frac{\phi_j(\hat{\eta}_j)}{\hat{\sigma}_{\phi_j 	(\hat{\eta}_j)}^2}\bigg/\sum_{j=1}^{k}\frac{1}{\hat{\sigma}_{\phi_j (\hat{\eta}_j)}^2},
\end{equation}
where, $\hat{\sigma}_{\phi_j (\hat{\eta}_j)}^2$ is the asymptotic variance of an unbiased estimator $\phi_j(\hat{\eta}_j)$. For small sample sizes, the MLEs of $\phi(\eta)$ have finite sample bias. However, as evident from equation \eqref{2.6}, a closed-form expression for the MLE of $\eta$ cannot be explicitly obtained. Therefore, it is substantially more complicated to obtain an unbiased MLE. Alternatively, one can resort to bias correction, which for the Weibull parameters under progressive type-II censoring, the corresponding study has been carried out by \cite{MT}. To do this, the following quantities are required in addition to the entries from the Fisher's information matrix:
\begin{equation}
	\mathscr{I}_j^{rs(t)} = \mathbb{E}\bigg[\frac{\partial}{\partial\theta_t}\mathscr{I}_j^{rs}\bigg]\quad\&\quad \mathscr{I}_j^{rst} = \mathbb{E}\bigg[\frac{\partial^3}{\partial\theta_r\partial\theta_s\partial\theta_t}\ell_j(\eta,\lambda_j)\bigg],\quad \theta = \eta,\lambda_j,~\&~r,s,t=1,2. \nonumber
\end{equation}
These quantities have been computed using special integrals of the form \eqref{2.7} \cite{ISG}. For instance, the quantities $\mathscr{I}_j^{rs(t)}$ are obtained as follows:
\begin{align}
	\mathscr{I}_j^{11(1)} &= -\frac{2}{\eta}\mathscr{I}_j^{11}, \quad
	\mathscr{I}_j^{11(2)} = -\frac{2}{\eta^2 	\lambda_j}\sum_{r_j=0}^{m_j-1}C_{r_j}\sum_{l=0}^{r_j} \frac{D_l^{(r_j)}}{R_l^{(r_j)}}\left[1-\left(\gamma + \log \lambda_j + \log R_l^{(r_j)}\right)\right], \nonumber \\
	\mathscr{I}_j^{12(1)} &= \mathscr{I}_j^{21(1)} = -\frac{1}{\eta}\mathscr{I}_j^{12}, \quad
	\mathscr{I}_j^{12(2)} = \mathscr{I}_j^{21(2)} = -\frac{1}{\lambda_j}\left[\mathscr{I}_j^{12} 	+ \frac{m_j}{\eta\lambda_j}\right], \nonumber \\
	\mathscr{I}_j^{22(1)} &= 0, \quad 		\mathscr{I}_j^{22(2)} = -\frac{2m_j}{\lambda_j^3}. \nonumber 
\end{align}
Using equation \eqref{2.7} the following two specific integrals are evaluated as:

\begin{align}
	I(1,2, R_l^{(r_j)},\eta,\lambda_j) =& 	\frac{1}{\eta^2\lambda_jR_l^{(r_j)}}\left[\left(1-\gamma-\log R_l^{(r_j)} - \log \lambda_j\right)^2 + \frac{\pi^2}{6}-1\right], \nonumber \\
	I(1,3, R_l^{(r_j)},\eta,\lambda_j) =& 	\frac{1}{\eta^3\lambda_jR_l^{(r_j)}}\bigg[\left(1-\gamma-\log R_l^{(r_j)} - \log \lambda_j\right)^3 + 3\left(\frac{\pi^2}{6}-1\right)\nonumber \\
	& \big(1-\gamma-\log R_l^{(r_j)} -\log \lambda_j\big)-2(\zeta(3)-1)\bigg], \nonumber
\end{align}
where $\zeta(3)\approx 1.2020569$. Then, the third derivative quantities $\mathscr{I}_j^{rst}$ are:
\begin{align}
	\mathscr{I}_j^{111} &= \frac{2m_j}{\eta^3} - \lambda_j 	\sum_{r_j=0}^{m_j-1}C_{r_j}\sum_{l=0}^{r_j}{D_l^{(r_j)}}({R_{(l+1)j}}+1)I(1,3, R_l^{(r_j)},\eta,\lambda_j), \nonumber \\
	\mathscr{I}_j^{112} &= \mathscr{I}_j^{121} = \mathscr{I}_j^{211} = -\sum_{r_j=0}^{m_j-1}C_{r_j}\sum_{l=0}^{r_j}{D_l^{(r_j)}}({R_{(l+1)j}}+1)I(1,2, R_l^{(r_j)},\eta,\lambda_j), \nonumber \\
	\mathscr{I}_j^{122} &= \mathscr{I}_j^{221} = \mathscr{I}_j^{212} = 0,\quad  \mathscr{I}_j^{122} = \frac{2m_j}{\lambda_j^3}. \nonumber
\end{align}
Consider the $2\times4$ matrix $\mathscr{A}$ defined as follows:
\begin{equation}
	\mathscr{A} = \begin{bmatrix}
		\mathscr{A}^{11(1)} & \mathscr{A}^{12(1)} & \mathscr{A}^{11(2)} & \mathscr{A}^{12(2)} \\
		\mathscr{A}^{21(1)} & \mathscr{A}^{22(1)} & \mathscr{A}^{21(2)} & \mathscr{A}^{22(2)} \\
	\end{bmatrix}, \nonumber
\end{equation} 
such that $\mathscr{A}^{rs(t)} = -\mathscr{I}_j^{rs(t)}-0.5\mathscr{I}_j^{rst}$. A first order 	bias-corrected MLE for $(\eta,\lambda_j)^T$ is given by
\begin{equation}
	\begin{bmatrix}
		\tilde{\eta} \\ \tilde{\lambda}_j
	\end{bmatrix} = \begin{bmatrix}
		\hat{\eta} \\ \hat{\lambda}_j
	\end{bmatrix} - \widehat{V}_j \hat{\mathscr{A}} \text{Vec}(\widehat{V}_j), 	\tag{2.9}\label{2.9}
\end{equation}
where, $\text{Vec}(\widehat{V}_j)=(\widehat{V}^{(11)}_j,\widehat{V}^{(12)}_j,\widehat{V}^{(21)}_j,\widehat{V}^{(22)}_j)$ is the row-vectorization formed by combining the rows of $\hat{V}_j$. The bias correction method by \cite{MT} is limited only for the two parameters. For functions of parameters $\Phi(\eta,\lambda_j)$, the bias correction is given by the following Theorem \ref{Th-2.2}:
\begin{theorem}\label{Th-2.2}
	The second order bias-corrected estimator of a smooth real-valued function $\Phi(.)$ of some parameter vector $\boldsymbol{\theta}$ is given by
	\begin{equation}
		\Phi(\tilde{\boldsymbol{\theta}}) = \Phi(\hat{\boldsymbol{\theta}}) - \nabla 	\Phi(\hat{\boldsymbol{\theta}})[\widehat{V} \hat{\mathscr{A}} \text{Vec}(\widehat{V})] -\frac{1}{2}\text{\normalfont tr}[H(\hat{\boldsymbol{\theta}})\widehat{V}], \tag{2.10}\label{2.10}
	\end{equation}
	where, $H(\hat{\boldsymbol{\theta}})$ is the Hessian matrix of $\Phi(\hat{\boldsymbol{\theta}})$.
	\begin{proof}
		The Taylor series expansion of $\Phi(\hat{\boldsymbol{\theta}})$ about the true value of $\Phi(\boldsymbol{\theta})$ is given by
		\begin{equation}
			\Phi(\hat{\boldsymbol{\theta}}) = \Phi(\boldsymbol{\theta}) + \nabla\Phi(\boldsymbol{\theta})(\hat{\boldsymbol{\theta}}-\boldsymbol{\theta})+\frac{1}{2}(\hat{\boldsymbol{\theta}}-\boldsymbol{\theta})^\top H(\boldsymbol{\theta}) (\hat{\boldsymbol{\theta}}-\boldsymbol{\theta}) + ... \tag{2.11}\label{2.11}
		\end{equation}
		Let $\tilde{\boldsymbol{\theta}}$ be the bias-corrected MLE of the form \eqref{2.9} such that $\tilde{\boldsymbol{\theta}}= \hat{\boldsymbol{\theta}} - \text{B}(\hat{\boldsymbol{\theta}})$, where $\text{B}(\hat{\boldsymbol{\theta}})=\widehat{V} \hat{\mathscr{A}} \text{Vec}(\widehat{V})$ is the first order bias correction. Note that $\mathbb{E}(\hat{\boldsymbol{\theta}}-\boldsymbol{\theta})=\text{B}(\hat{\boldsymbol{\theta}})$. Next, on taking expectation of the quadratic form in Equation \eqref{2.11},
		\begin{equation}
			\mathbb{E}[(\hat{\boldsymbol{\theta}}-\boldsymbol{\theta})^\top H(\boldsymbol{\theta})(\hat{\boldsymbol{\theta}}-\boldsymbol{\theta})] = \text{tr}[H(\boldsymbol{\theta})\hat{V}] + O(n^{-2}). \nonumber
		\end{equation}
		Therefore, taking expectation of Equation \eqref{2.11}, the bias is obtained as:
		\begin{equation}
			\mathbb{E}[\hat{\Phi}(\boldsymbol{\theta}) - \Phi(\boldsymbol{\theta})] =  \nabla\Phi(\boldsymbol{\theta})\text{B}(\hat{\boldsymbol{\theta}}) + \text{tr}[H(\boldsymbol{\theta})\hat{V}] + O(n^{-2}). \nonumber
		\end{equation}
		Furthermore, by the multivariate delta-method, $\nabla\Phi(\hat{\boldsymbol{\theta}})\text{B}(\hat{\boldsymbol{\theta}}) - \nabla\Phi(\boldsymbol{\theta})\text{B}(\hat{\boldsymbol{\theta}}) = O_p(n^{-\frac{3}{2}})$, which is asymptotically negligible. Similarly for $H(\boldsymbol{\theta})$. Therefore the MLE $\hat{\boldsymbol{\theta}}$ can be plugged-in the quantities $\nabla\Phi(\boldsymbol{\theta})$ and $H(\boldsymbol{\theta})$. Hence, the theorem follows.
	\end{proof}
\end{theorem}
Let $\Phi(\boldsymbol{\theta}) = \phi_j(\eta_j)$. The bias correction of the MLE would also call for adjustment of the asymptotic variance estimate $\hat{\sigma}_{\phi_j (\hat{\eta}_j)}^2$. However, the following Theorem \ref{Th-2.3} suggests otherwise. 
\begin{theorem} \label{Th-2.3}
	The leading first-order asymptotic variance of the bias-corrected MLE  is identical to that of the ordinary MLE.
	\begin{proof}
		Recall that $\tilde{\boldsymbol{\theta}}= \hat{\boldsymbol{\theta}} - \text{B}(\hat{\boldsymbol{\theta}})$. Then, the variance of $\tilde{\boldsymbol{\theta}}$ is
		\begin{align}
			\nonumber
			\tilde{\sigma}_{\tilde{\boldsymbol{\theta}}}^2 = \widehat{\text{Var}}(\hat{\boldsymbol{\theta}} - \text{B}(\hat{\boldsymbol{\theta}})) = \hat{\sigma}_{\hat{\boldsymbol{\theta}}}^2 + &\hat{\sigma}_{\text{B}(\hat{\boldsymbol{\theta}})}^2 - 2\text{Cov}(\hat{\boldsymbol{\theta}},\text{B}(\hat{\boldsymbol{\theta}})) = \hat{\sigma}_{\hat{\boldsymbol{\theta}}}^2 + O(n^{-2}).\\
			\implies \tilde{\sigma}_{\tilde{\boldsymbol{\theta}}}^2 &\approx \hat{\sigma}_{\hat{\boldsymbol{\theta}}}^2.\nonumber
		\end{align}
		The same logical justification can be applied for the variance of $\Phi(\tilde{\boldsymbol{\theta}})$, that is, $\tilde{\sigma}_{\Phi(\tilde{\boldsymbol{\theta}})}^2 \approx \hat{\sigma}_{\Phi(\hat{\boldsymbol{\theta}})}^2 = [\nabla\Phi(\boldsymbol{\theta})]\mathscr{I}^{-1}(\boldsymbol{\theta})[\nabla\Phi(\boldsymbol{\theta})]^\top_{\boldsymbol{\theta}=\hat{\boldsymbol{\theta}}}$.
	\end{proof}
\end{theorem}
To construct the large sample interval of $\phi(\eta)$, the following property below is used:
\begin{theorem}
	Let $\phi(\tilde{\eta})$ be the pooled-in estimator of $\phi(\eta)$. If each $\phi_j(\tilde{\eta}_j)\sim N(\phi(\eta),\tilde{\sigma}_{\phi_j(\tilde{\eta}_j)}^2)$, then $\phi(\tilde{\eta})\sim N(\phi(\eta),\tilde{\sigma}_{\phi (\tilde{\eta})}^2)$, where $\tilde{\sigma}_{\phi (\tilde{\eta})}^2 = \left({\sum_{j=1}^{k}1/{\tilde{\sigma}_{\phi_j (\tilde{\eta}_j)}^2}}\right)^{-1}$.
	\begin{proof}
		\normalfont Let $a_j = 1/\tilde{\sigma}_{\phi_j (\tilde{\eta}_j)}^2$. Then, equation \eqref{2.8} can be rewritten as:
		\begin{equation} \label{2.12}\tag{2.12}
			\phi(\tilde{\eta}) = \sum_{j=1}^{k} a_j \phi_j (\tilde{\eta}_j) / \sum_{j=1}^{k} a_j.
		\end{equation}
		It is a well-known fact that a linear combination of independent normal variables is normal \cite{GS}. This implies that the weighted average of independent normal variables, such as in this case, the quantity $\phi(\tilde{\eta})$ follows normal distribution with the following moment generating function:
		\begin{equation} \nonumber
			M_{\phi(\tilde{\eta})} (t) = \exp\bigg[\big(\sum_{j=1}^{k} {a_j {\phi}(\eta)}/{\sum_{j=1}^{k} a_j}\big) t + \frac{1}{2} \bigg(\frac{1}{\sum_{j=1}^{k} a_j} \bigg) t^2 \bigg],
		\end{equation}
		and hence, the theorem follows.
	\end{proof}
\end{theorem}
Therefore, a $100(1-\alpha)\%$ large sample interval of $\phi(\eta)$ is given by:
\begin{equation} \label{2.13}\tag{2.13}
	\textbf{I}_{LS} = \left[\phi(\tilde{\eta}) \mp z_{1-\frac{\alpha}{2}} 	\sqrt{\left({\sum_{j=1}^{k} \frac{1}{\tilde{\sigma}_{\phi_j (\tilde{\eta}_j)}^2}}\right)^{-1}}\right],
\end{equation}
where, $z_{1-\frac{\alpha}{2}}$ denotes the $(1-\frac{\alpha}{2})^\text{th}$ quantile of the standard normal distribution and $\tilde{\sigma}_{\phi_j (\tilde{\eta}_j)}^2$ is the leading $O(n^{-1})$ asymptotic variance of $\phi_j (\tilde{\eta}_j)$.
\subsection{Method of Variance Estimates Recovery}
The method of variance estimates recovery (MOVER) was introduced by \cite{GYZ} and is generally utilized to construct confidence intervals for a linear combination of parameters. Through this method, the variance estimate is obtained at the end-points of the $z$-type intervals. Some authors have also proposed using $t$-type intervals, for example in  \cite{KK} for a weighted lognormal mean.\\
\indent Let $\left[l_j, u_j\right]$ represent a known interval for $\phi(\eta)$ from the $j^\text{th}$ sample. The $z$-type intervals are of the form $\left(\max\left\{0,\phi_j\left(\hat{\eta}_j\right)-z_{1-\frac{\alpha}{2}}\sqrt{\hat{\sigma}_{\phi_j(\hat{\eta}_j)}^2}\right\},\phi_j(\hat{\eta}_j)+z_{1-\frac{\alpha}{2}}\sqrt{\hat{\sigma}_{\phi_j(\hat{\eta}_j)}^2}\right)$. The restriction on the lower bound brings about a loss of precision. Therefore in this paper, the log-Wald type interval is considered using the asymptotic properties of log-transformed bias-corrected MLEs. Therefore, consider the following interval for $\phi(\eta)$ using the $j^{\text{th}}$ sample:
\begin{equation} \label{2.14}\tag{2.14}
	[l_j,u_j] = \left[\phi_j\left(\tilde{\eta}_j\right)\exp\left(-z_{1-\frac{\alpha}{2}} \frac{\sqrt{\tilde{\sigma}_{\phi_j\left(\tilde{\eta}_j\right)}^2}}{\phi_j\left(\tilde{\eta}_j\right)}\right), ~\phi_j\left(\tilde{\eta}_j\right)\exp\left(z_{1-\frac{\alpha}{2}} \frac{\sqrt{\tilde{\sigma}_{\phi_j\left(\tilde{\eta}_j\right)}^2}}{\phi_j\left(\tilde{\eta}_j\right)}\right)\right].
\end{equation}
The MOVER is applied for computation of the variance estimates at $\phi(\eta) = l_j$ and $\phi(\eta) = u_j$, which are given as follows:
\begin{equation} \nonumber
	\tilde{\sigma}^2_{\phi_j(\tilde{\eta}_j),l_j} = \frac{\left[\phi_j(\tilde{\eta}_j) 	-l_j\right]^2}{z_{1-\frac{\alpha}{2}}^2} \quad \& \quad \tilde{\sigma}^2_{\phi_j(\tilde{\eta}_j),u_j} = \frac{\left[u_j - \phi_j(\tilde{\eta}_j)\right]^2}{z_{1-\frac{\alpha}{2}}^2}.
\end{equation}
On combining the two estimates, the variance estimate of $\phi_j(\tilde{\eta}_j)$ at $l_j$ and $u_j$ are given by
\begin{align}
	\tilde{\sigma}^{*2}_{\phi_j(\tilde{\eta}_j)} &= 	\frac{1}{2z_{1-\frac{\alpha}{2}}^2}\left\{\left[\phi_j(\tilde{\eta}_j) -l_j\right]^2 + \left[u_j - \phi_j\left(\tilde{\eta}_j\right)\right]^2\right\} \nonumber \\
	&= 	\frac{\left[\phi_j(\tilde{\eta}_j)\right]^2}{2z_{1-\frac{\alpha}{2}}^2}\left\{\left[1-\exp\left(-z_{1-\frac{\alpha}{2}} \frac{\sqrt{\tilde{\sigma}_{\phi_j(\tilde{\eta}_j)}^2}}{\phi_j(\tilde{\eta}_j)}\right)\right]^2 + \left[\exp\left(z_{1-\frac{\alpha}{2}} \frac{\sqrt{\tilde{\sigma}_{\phi_j(\tilde{\eta}_j)}^2}}{\phi_j(\tilde{\eta}_j)}\right)-1\right]^2\right\}. \label{2.15}\tag{2.15}
\end{align}
Considering the variance estimate as $\tilde{\sigma}^{*2}_{\phi_j(\tilde{\eta}_j)}$, the pooled-in estimator of $\phi(\eta)$ is given by
\begin{equation} \nonumber
	\phi^*(\tilde{\eta}) = \sum_{j=1}^{k} 	\frac{\phi_j(\tilde{\eta}_j)}{\tilde{\sigma}^{*2}_{\phi_j(\tilde{\eta}_j)}}\bigg/\sum_{j=1}^{k}\frac{1}{\tilde{\sigma}^{*2}_{\phi_j(\tilde{\eta}_j)}}.
\end{equation}
Therefore, the bounds of the $(1-{\alpha})100\%$ approximate confidence interval for $\phi(\eta)$ due to MOVER are given by $\textbf{I}_{MOVER} = [L, U]$, where
\begin{equation} \label{2.16}\tag{2.16}
	L = \left[\phi(\tilde{\eta}) - z_{1-\frac{\alpha}{2}}\sqrt{\frac{1}{\sum_{j=1}^{k}1/\tilde{\sigma}^2_{\phi_j(\tilde{\eta}_j),l_j}}}\right] \quad \& \quad U= \left[\phi(\tilde{\eta}) + z_{1-\frac{\alpha}{2}}\sqrt{\frac{1}{\sum_{j=1}^{k}1/\tilde{\sigma}^2_{\phi_j(\tilde{\eta}_j),u_j}}}\right].
\end{equation}
\subsection{Generalized Pivotal Method}
The frequentist methods discussed earlier have heavy dependency on asymptotic normality properties. Generalized confidence intervals ($\textbf{I}_{GC}$) are a modern extension of classical confidence interval theory introduced by \cite{SW}, who developed the concept of generalized pivotal quantities (GPQs). Let $\boldsymbol{x}$ be the vector of observed values of the random sample $\boldsymbol{X}$ that depends on some random vector $(\boldsymbol{\theta},\boldsymbol{\delta})$, where $\boldsymbol{\theta}$ is the parameter of interest and $\boldsymbol{\delta}$ is the vector of nuisance parameters. A function $\rho\equiv\rho(\boldsymbol{X};\boldsymbol{x},\boldsymbol{\theta},\boldsymbol{\delta})$ is called a GPQ if its probability distribution does not contain any unknown parameter and its observed value is independent of any nuisance parameters. \cite{BXW} proposed an approach for obtaining the GPQ of $\phi(\eta)$, for a single progressive type-II censored sample. Therefore, for each $j=1,2,...k,$ consider the following properties: \\
1. Let $X_{ij:m_j:m_j}$, where $i=1,2,...,m_j$ and $j=1,2,...,k$ be $m_j$ ordered statistics from the Weibull($\eta,\lambda_j$) distribution. Then, $Y_{ij:m_j:m_j} = \lambda_j x_{ij:m_j:m_j}^{\eta} \sim \text{Exponential}(1)$. \\
2. Let $D_{1j:m_j:m_j} = n_j Y_{1j:m_j:m_j}$, and $D_{ij:m_j:m_j} = [n_j - \sum_{l=1}^{i-1}(1+R_{lj})](Y_{ij:m_j:m_j} - Y_{(i-1)j:m_j:m_j})$ for all $i=2,...,m_j$. Then, the distance statistics $D_{ij:m_j:m_j}$ are ordered standard exponential variables with unit rate for all $i=1,2,...,m_j$ \cite{RV}. \\
3. If $Z_i^{(j)} = \sum\limits_{l=1}^{i} D_{lj:m_j}$, $i=1,2,...,m_j$ and $U_{(i)}^{(j)} = Z_i^{(j)}/Z_{m_j}^{(j)}$, for $i=1,2,...,m_j-1$, then $U_{(1)}^{(j)} < U_{(2)}^{(j)} < ... < U_{(m_j - 1)}^{(j)}$ are ordered statistics with each $U_{(i)}^{(j)} \sim U(0,1)$, $i=1,2,...,m_j-1$. It can be further shown that
\begin{equation} \nonumber
	Z_i^{(j)} = \sum_{l=1}^{i} \left(1+R_{lj}\right)Y_{lj:m_j:m_j} + \left[n_j - 	\sum_{l=1}^{i}\left(1 + R_{lj}\right)\right]Y_{ij:m_j:m_j};\quad i=1,2,...,m_j.
\end{equation}
Consider the function $\mathscr{V}_j(\eta) = \sum_{i=1}^{m_j}(-2\log U_{(i)}^{(j)})$. Then it follows that $\forall~j=1, 2,...,k$:
\begin{equation}
	\label{2.17}\tag{2.17}
	\mathscr{V}_j(\eta) = 2\sum_{i=1}^{m_j-1}\log 	\left[\frac{\sum_{l=1}^{m_j}\left(1+R_{lk}\right)X_{lj:m_j:m_j}^{\eta}}{\sum_{l=1}^{i}\left(1+R_{lj}\right)X_{lj:m_j:m_j}^{\eta} + \left\{n_j - \sum_{l=1}^{i}\left(1 + R_{lj}\right)\right\} X_{i:m_j:m_j}^{\eta}}\right]. 
\end{equation}
According to \cite{VKR}, if for any $j=1,2,...,k$, $U_{i}^{(j)}\sim U(0,1)$, then $-2\log U_{i}^{(j)}\sim \chi^2_{(2)}$. By commutativity of addition,
\begin{equation}
	\label{2.18}\tag{2.18}
	\mathscr{V}_j(\eta)= \sum_{i=1}^{m_j} (-2 \log U_{(i)}^{(j)}) =  \sum_{i=1}^{m_j} (-2 \log 	U_{i}^{(j)}) \sim \chi^2_{2(m_j-1)};\quad j=1,2,...,k.
\end{equation}
It can be shown that $\mathscr{V}_j(\eta) > 0$ and is strictly monotonic for all $\eta>0$. Therefore, $\mathscr{V}_j$ has an inverse which is denoted by $\mathscr{V}_j^{-1}$. Its expression is free of the nuisance parameter $\lambda_j$, and its distribution depends only on $m_j$ and not on any unknown parameters. Let $\rho_{\phi(\eta)}$ be the GPQ of $\phi(\eta)$. The GPQ of $\phi(\eta)=\eta$ based on the $j^\text{th}$ sample is denoted by $\rho_{\phi(\eta), j}$ and is given by the unique solution of $\eta_j=\mathscr{V}_j^{-1}(\tau)$, where $\tau \sim \chi^2_{2(m_j-1)}$. For $\phi(\eta) = \text{C}_\eta$, the GPQ based on the $j^\text{th}$ sample is given by
\begin{equation} \label{2.19}\tag{2.19}
	\rho_{\text{C}_\eta,j} = 	\sqrt{\frac{\Gamma\left(1+\frac{2}{\mathscr{V}_j^{-1}(\tau)}\right)}{\Gamma^2\left(1+\frac{1}{\mathscr{V}_j^{-1}(\tau)}\right)} - 1}.
\end{equation}
Let $\boldsymbol{\rho}_j = \left\{\rho_{\phi(\eta), j}^{(1)},\rho_{\phi(\eta), j}^{(2)},...,\rho_{\phi(\eta), j}^{(B)}\right\}$ be the sample obtained by generating $\rho_{\phi(\eta), j}$ for a fixed $B$ times. The pivotal variance of the pivotal distribution of $\rho_{\phi(\eta), j}$ can be approximated by the sample variance as follows:
\begin{equation} \label{2.20}\tag{2.20}
	\hat{\sigma}^2(\rho_{\phi(\eta),j}) = 	\frac{1}{B-1}\sum_{b=1}^{B}\left[\rho_{\phi(\eta),j}^{(b)}-\overline{\rho_{\phi(\eta),j}}\right]^2,
\end{equation}
where, $\overline{\rho_{\phi(\eta),j}}$ is the sample mean of $\boldsymbol{\rho}_j$. It can be observed that $\hat{\sigma}(\rho_{\phi(\eta), j})$ is a function of $\rho_{\phi(\eta), j}^{(b)}$ only, and is also independent of unknown parameters. Therefore, the pooled-in GPQ for $\phi(\eta)$ can be approximated as follows:
\begin{equation} \label{2.21}\tag{2.21}
	\rho_{\phi(\eta)} = \sum_{j=1}^{k} \frac{\rho_{\phi(\eta), 	j}}{\hat{\sigma}^2(\rho_{\phi(\eta),j})}\bigg/\sum_{j=1}^{k}\frac{1}{\hat{\sigma}^2(\rho_{\phi(\eta),j})},
\end{equation}
Therefore, the 100$(1-\alpha)\%$ two-sided confidence intervals for $\phi(\eta)$ based on the GPQs is
\begin{equation}
	\tag{2.22}\label{2.22}
	\textbf{I}_{GC} = \bigg[\rho_{\phi(\eta)}\bigg(\frac{\alpha}{2}\bigg), \rho_{\phi 	(\eta)}\bigg(1-\frac{\alpha}{2}\bigg)\bigg],
\end{equation}
where, $\rho_{\phi (\eta)}(1-\frac{\alpha}{2})$ represents the $100(1-\frac{\alpha}{2})^\text{th}$ upper percentile of the pivotal distribution of $\rho_{\phi(\eta)}$. The flow of this method is summarized in the Algorithm \ref{Al-GCI}.
\begin{algorithm}[H]
	\caption{(Generalized Pivotal Method)}\label{Al-GCI}
	\begin{algorithmic}[1]
		\For{$j=1,2,...,k$}
		\For{$b=1,2,...,B$}
		\State Generate $\tau \sim \chi^2_{2(m_j-1)}$.
		\State Solve for $\eta_j$ using equation \eqref{2.18}.
		\State Determine $\rho_{\phi(\eta), j}^{(b)}$.
		\EndFor
		\State Compute $\hat{\sigma}(\rho_{\phi(\eta),j})$ using \eqref{2.20}
		\EndFor
		\For{$b=1,2,...,B$}
		\State Calculate $\rho_{\phi(\eta)}^{(b)}$ using equation \eqref{2.21}
		\EndFor
		\State Sort $\rho_{\phi(\eta)}^{(b)}$, $b=1,2,...,B$ in increasing order:
		\begin{align*} 
			\rho_{\phi(\eta), (1)} \leq \rho_{\phi(\eta), (2)}& \leq \hdots \leq \rho_{\phi(\eta), (B)}.
		\end{align*} 
		\State Determine $\textbf{I}_{GC}$ using equation \eqref{2.22}.
	\end{algorithmic}
\end{algorithm}
\subsection{Bayesian Methods}
In this subsection, the framework of Bayesian interval estimation is discussed, focusing on its two primary types--the equal-tailed credible and highest posterior density intervals. The previous frequentist approaches are used in quantifying the confidence intervals by regarding the parameters as unknown but fixed quantities. However, Bayesian inference considers a probabilistic framework for the parameters by treating them as random variables with some prior knowledge, which is quantified by a probability distribution known as the prior distribution. Then, upon integrating the prior distribution with the observed data using Bayes' theorem, the posterior distribution is obtained. \\
\indent Bayesian interval estimation reflects the actual probability of the parameter lying within specific bounds, given the observed data.  That is, the Bayesian credible intervals are defined as the regions of the parameter space that contain a specified probability mass of its posterior distribution. For simple conjugate models, the exact form of the intervals can usually be determined easily. However, for nonlinear, high-dimensional, and censored models, the Markov Chain Monte Carlo (MCMC) algorithms are preferred to compute empirical quantiles or densities from the samples. The MCMC algorithms are a cornerstone of modern Bayesian inference, particularly for computing posterior distributions and the associated interval estimates. The $100(1-\alpha)\%$ equal-tailed credible interval ($\textbf{I}_{BC}$) for some parameter $\phi$ is obtained from these empirical quantiles of the posterior samples, and is given by $\textbf{I}_{BC} = [\phi_{(\frac{\alpha}{2})}, \phi_{(1 - \frac{\alpha}{2})}]$, where $\phi_{(\frac{\alpha}{2})}$ denotes the lower $100{(\frac{\alpha}{2})}^\text{th}$ percentile of the empirical posterior distribution of $\phi(\eta)$. Furthermore, the highest posterior density interval ($\textbf{I}_{HPD}$) is the shortest interval among all those intervals that contain posterior probability $(1-\alpha)$. If the ordered posterior samples are $\{\phi_{(r)}\}_{r=1}^{\mathscr{R}}$, then the $100(1-\alpha)\%$ $\textbf{I}_{HPD}$ of ${\phi}$ is  $[\phi_{(r)}, \phi_{(r^*+\mathscr{R}\{1-\alpha\})}]$, where $r^*$ is given by:
\begin{equation} \tag{2.23}\label{2.23} 
	r^* = \arg\min_{1\leq r \leq \mathscr{R}\alpha} [\phi_{(r + \mathscr{R}\{1-\alpha\})}- 	\phi_{(r)}].
\end{equation}
Although the populations share the same shape parameter $\eta$, the posterior distribution under a fully joint Bayesian model involves $\eta$ and all $k$ nuisance rate parameters. The resulting posterior distribution may require high-dimensional MCMC sampling, especially when $k$ is large. To reduce computational complexity, the common shape parameter is estimated separately from each sample. The gamma prior is highly flexible and computationally convenient for parameters with positive real support. Assume that $\eta_j$ and $\lambda_j$ follow independent prior gamma distributions with rate parametrization, that is, $\pi_{\eta_j}(\eta_j) \sim \text{Gamma} (a_j, b_j)$ and $\pi_{\lambda_j}(\lambda_j) \sim \text{Gamma} (c_j, d_j)$; where, $a_j, b_j, c_j, d_j > 0$ are the prior hyperparameters and $j= 1, 2,...,k$. The resulting joint posterior distribution for each $j=1,2,...,k$ is proportionally given by:
\begin{align} \nonumber
	\pi (\eta_j, \lambda_j) \propto&	\mathscr{L}(\eta_j,\lambda_j)\pi_{\eta_j}(\eta_j)\pi_{\lambda_j}(\lambda_j)\\
	=& \eta_j^{a_j+m_j- 1} \lambda_j^{c_j+m_j- 1} \exp[(\eta_j - 1)\sum_{i=1}^{m_j}\log 	x_{i:m_j:n_j} - b_j \eta_j - \lambda_j\{d_j + \sum_{i=1}^{m_j} (1+ R_{ij}) 	x_{i:m_j:n_j}^{\eta_j}\}]. \nonumber
\end{align}
This gives:
\begin{align} \tag{2.24}\label{2.24}
	\lambda_j~|~\eta_j \sim \text{Gamma}(c_j + m_j, d_j& + \sum_{i=1}^{m_j} (1+R_{ij}) 	x_{i:m_j:n_j}^{\eta_j}), \\  \tag{2.25}\label{2.25}
	\eta_j~|~\lambda_j \sim \eta_j^{a_j+m_j- 1}\exp[(\eta_j - 1)\sum_{i=1}^{m_j}\log 	x_{i:m_j:n_j} &- b_j \eta_j - \lambda_j\sum_{i=1}^{m_j} (1+R_{ij}) x_{i:m_j:n_j}^{\eta_j}].
\end{align}
It is noticed that the conditional posterior distribution of $\eta_j$ cannot be expressed in any of the well-known canonical distributions. To determine the Bayesian intervals for $\phi(\eta)$, the Monte Carlo algorithm by hybrid slice-within-Gibbs sampling is applied to generate an ergodic Markov chain whose stationary distribution is the target posterior density. The core idea behind the slice sampling algorithm \cite{RMN} is to uniformly sample $\eta_j$ from the area under the density curve by alternately drawing the auxiliary height $y \sim U(0, \pi (\eta_j, \lambda_j))$. Then the new state is uniformly sampled from the horizontal region $\{\eta_j: y < \pi (\eta_j, \lambda_j)\}$ called the ``slice". Suppose at an iterative step $r$, the Gibbs sampler has initially generated $\lambda_j^{(r)}$ such that $\pi(\eta^{(r-1)}, \lambda_j^{(r)}) > 0$. A random vertical height $y$ is sampled uniformly from the interval $[0, \pi(\eta_j^{(r-1)}, \lambda_j^{(r)})]$. Then, two stages are immediately followed: the first stage, which is known as the stepping out algorithm, requires that an initial interval of some width $w$ be placed around $\eta_j^{(r-1)}$, and is expanded or ``stepped-out" until the exterior of the slice is reached. The budget of the stepping-out algorithm can be constrained by limiting the size of the slice to $wM$, for some $M\in \mathbb{Z}$. This results in an auxiliary interval $\mathscr{I} = (\mathscr{I}_L, \mathscr{I}_R)$ which contains $\eta_j^{(r-1)}$ along with as much of the slice as possible. The second stage algorithm shrinks this interval until a point $\eta_j^{(r)}$ is obtained such that $y < \pi(\eta_j^{(r)}, \lambda_j^{(r)})$.  For the $j^\text{th}$ progressive type-II censored sample, the inverse posterior variance pooled-in Bayesian estimator of $\phi(\eta)$ based on the Bayesian method at the $r^\text{th}$ iteration is:
\begin{equation}  \tag{2.26}\label{2.26}
	\phi(\check{\eta}^{(r)}) = \sum_{j=1}^{k} 	\frac{\{\phi_j(\check{\eta}^{(r)})\}}{\{\check{\sigma}_{\phi_j (\check{\eta}^{(r)})}^2\}}\bigg/\sum_{j=1}^{k}\frac{1}{\{\check{\sigma}_{\phi_j (\check{\eta}^{(r)})}^2\}},
\end{equation}
\cite{MLO} have utilized the Delta method to obtain the variance estimate of the common mean. However, the Bayesian inference is built on the posterior distribution rather than on asymptotic approximations. Therefore, it is more meaningful to use the empirical estimate $\{\check{\sigma}_{\phi_j (\check{\eta}_j)}^2\}^{(r)}$ of the posterior variance obtained from the MCMC samples themselves:
\begin{equation} \tag{2.27}\label{2.27}
	\check{\sigma}_{\phi_j (\check{\eta})}^2 = 	\frac{1}{\mathscr{R}-1}\sum_{r=1}^{\mathscr{R}}\left[\{\phi_j(\check{\eta}^{(r)})\}-\overline{\{\phi_j(\check{\eta})\}}\right]^2
\end{equation}
The flow of this framework is provided in Algorithm \ref{Al-Bayesian}.
\begin{algorithm}[t!]
	\caption{(Bayesian Intervals - Slice-within-Gibbs Sampling)}\label{Al-Bayesian}
	\begin{algorithmic}[1]
		\For{$j=1,2,...,k$}
		\State Fix $\mathscr{R}$, $w$, $M$ and let $r=1$, and initialize $\eta_j^{(0)}$ and $\lambda_j^{(0)}$.
		\For{$r = 1,2,...,\mathscr{R}$} 
		\State Draw $\lambda_j^{(r)}$ using \eqref{2.24}.
		\State Draw $y \sim {U}[0, \pi (\eta_j^{(r-1)},\lambda_j^{(r)})]$.
		\State Draw $u\sim U(0,1)$ and compute $\mathscr{I}_L = \eta_j^{(r-1)}-wu$, $\mathscr{I}_R = \mathscr{I}_L+w$.
		\State Draw $z\sim U(0,1)$ and compute step-out budgets $B_l=\lfloor zM \rfloor$, $B_r = M-B_l-1$.
		\While{$y < \max \{0,\pi(\mathscr{I}_R,\lambda_j^{(r-1)})\}~\&~B_r > 0$}
		\State Update $\mathscr{I}_R <- \mathscr{I}_R + w$ and $B_r <- B_r - 1$.
		\EndWhile 
		\While{$y < \max \{0,\pi(\mathscr{I}_L,\lambda_j^{(r-1)})\}~\&~B_l > 0$}
		\State Update $\mathscr{I}_L <- \mathscr{I}_L - w$ and $B_l <- B_l - 1$.
		\EndWhile 
		\State Set $\delta=0$, $L^* = \mathscr{I}_L$ and $R^* = \mathscr{I}_R$.
		\While{$\delta=0$}
		\State Draw $u^*\sim U(0,1)$ and compute $\eta_j^{(r)} = L^* + u^*(R^*-L^*)$.
		\If{$y <\max\{0,\pi(\eta_j^{(r)},\lambda_j^{(r)})\}$} 
		\State Set $\delta=1$ and exit loop.
		\EndIf
		\If{$\eta_j^{(r)} < \eta_j^{(r-1)}$}
		\State $L^* = \eta_j^{r}$
		\Else
		\State $R^* = \eta_j^{(r)}$
		\EndIf
		\EndWhile 
		\State Compute $\phi_j(\eta_j^{(r)}) = \text{C}_{\eta_j}^{(r)}$.
		\State Determine $\check{\sigma}_{\phi_j (\check{\eta}_j)}^2$ using equation \eqref{2.27}.
		\EndFor
		\EndFor
		\For{$r=1,2,...,\mathscr{R}$}
		\State Compute $\phi(\check{\eta})^{(r)}$ using equation \eqref{2.26}.
		\EndFor 
		\State Sort $\phi(\check{\eta})^{(r)}$, $r=1,2,...,R$ in increasing order:
		\begin{align*} 
			\phi(\check{\eta})_{(1)} \leq \phi(\check{\eta})_{(2)}& \leq \hdots \leq \phi(\check{\eta})_{(\mathscr{R})}.
		\end{align*} 
		\State Determine the $\textbf{I}_{BC}$ and $\textbf{I}_{HPD}$ as discussed.
	\end{algorithmic}
\end{algorithm}
\section{Optimal Censoring Design}\label{sec4}
In lifetime data analysis, the phenomenon of censoring has usually been treated as a necessary inconvenience, but modern experimental design considers it as an active variable. For the case of progressive censoring, while the removals in some circumstances are beyond the control of the experimenter, the ability to achieve maximum precision is also a necessary factor for robust inference. Several practical cases, however, allow the experimenter to transform censoring from a data-loss problem to an optimized parameter. These optimal censoring designs for $k=1$ are usually achievable upon optimizing specific information-related criteria according to the model specification. For higher-dimensional populations, the information matrix becomes block diagonal, bringing about an intractable mathematical challenge. Therefore, in this section, a method based on the minimization of the variance estimate of the parameter $\phi(\eta)$ is considered. \\
\indent From Theorem 2.4, upon considering the independence conditions of the $k$ progressive type-II censored samples, the overall asymptotic variance of the pooled-in estimator $\phi(\tilde{\eta})$ is given by $\tilde{\sigma}_{\phi (\tilde{\eta})}^2 = \big[{\sum_{j=1}^{k}{\tilde{\sigma}_{\phi_j (\tilde{\eta}_j)}^{-2}}}\big]^{-1}$. Consider the objective function to be minimized:
\begin{equation} \tag{3.1}\label{3.1}
	J(\boldsymbol{R}_1,\boldsymbol{R}_2,...,\boldsymbol{R}_k) = \tilde{\sigma}_{\phi 	(\tilde{\eta})}^2~|~\boldsymbol{R}_1,\boldsymbol{R}_2,...,\boldsymbol{R}_k = \left(\sum_{j=1}^{k}\frac{1}{\tilde{\sigma}_{\phi_j (\tilde{\eta}_j)}^{2}}~|~\boldsymbol{R}_j\right)^{-1},
\end{equation}
where $\boldsymbol{R}_j = \{R_{1j},R_{2j},...,R_{m_jj}\}$. Alternatively, suppose for each $j=1, 2,...,k$, and fixed ${\phi_j(\tilde{\eta}_j)}$, the set $\boldsymbol{R}^*_j$ minimizes $\tilde{\sigma}_{\phi_j (\tilde{\eta}_j)}^2$. Then for any other set $\boldsymbol{R}_j$:
\begin{equation}
	\tag{3.2}\label{3.2}
	\tilde{\sigma}_{\phi_j (\tilde{\eta}_j)}^2~|~\boldsymbol{R}_j \geq
	\tilde{\sigma}_{\phi_j (\tilde{\eta}_j)}^2~|~\boldsymbol{R}^*_j =
	\min\limits_{\boldsymbol{R}_j}\tilde{\sigma}_{\phi_j (\tilde{\eta}_j)}^2.
\end{equation}
This implies that for all $j=1,2,...,k$,
\begin{equation} \nonumber
	\frac{1}{\tilde{\sigma}_{\phi_j (\tilde{\eta}_j)}^2}~|~\boldsymbol{R}_j \leq
	\frac{1}{\tilde{\sigma}_{\phi_j (\tilde{\eta}_j)}^2}~|~\boldsymbol{R}^*_j \implies 	\sum_{j=1}^{k} \left[\frac{1}{\tilde{\sigma}_{\phi_j (\tilde{\eta}_j)}^2}~|~\boldsymbol{R}_j\right] \leq \sum_{j=1}^{k}\left[ \frac{1}{\tilde{\sigma}_{\phi_j (\tilde{\eta}_j)}^2}~|~\boldsymbol{R}^*_j\right].
\end{equation}
Since $\forall$ $x>0$, the function $f(x) = \frac{1}{x}$ is strictly decreasing, therefore
\begin{align} \nonumber
	\left(\sum_{j=1}^{k}\left[\frac{1}{\tilde{\sigma}_{\phi_j (\tilde{\eta}_j)}^2}~|~\boldsymbol{R}_j\right]\right)^{-1} \geq& \left(\left[\sum_{j=1}^{k} \frac{1}{\tilde{\sigma}_{\phi_j (\tilde{\eta}_j)}^2}~|~\boldsymbol{R}^*_j\right]\right)^{-1}, \\
	\text{i.e.,}~ J(\boldsymbol{R}_1,\boldsymbol{R}_2,...,\boldsymbol{R}_k) \geq& \sum_{j=1}^{k} J(\boldsymbol{R}^*_1,\boldsymbol{R}^*_2,...,\boldsymbol{R}^*_k). \nonumber
\end{align}
which implies that the individual optimal censoring schemes of each sample that minimizes the corresponding asymptotic variance of $\phi_j(\tilde{\eta}_j)$ are all collectively optimal for $\phi(\tilde{\eta})$. Hence, the global optimal censoring plans $(\boldsymbol{R}^*_1,\boldsymbol{R}^*_2,...,\boldsymbol{R}^*_k)$ are obtained from the constrained minimization problem below:
\begin{equation}
	\tag{3.3}\label{3.3}
	(\boldsymbol{R}^*_1,\boldsymbol{R}^*_2,...,\boldsymbol{R}^*_k) = 	\arg\min\limits_{\boldsymbol{R}_1,\boldsymbol{R}_2,...,\boldsymbol{R}_k} \left(\sum_{j=1}^{k}\frac{1}{\tilde{\sigma}_{\tilde{\phi}_j (\eta_j)}^{2}}\right)^{-1}  ,
\end{equation}
subject to ${R}_{ij} \geq 0$ for each $i=1,2,...,m_j$, $j=1,2,...,k$ and $n_j = \sum\limits_{i=1}^{m_j} (1+R_{ij})$.
\begin{figure}[b!]
	\centering
	\includegraphics[width=0.95\linewidth, height=0.375\linewidth]{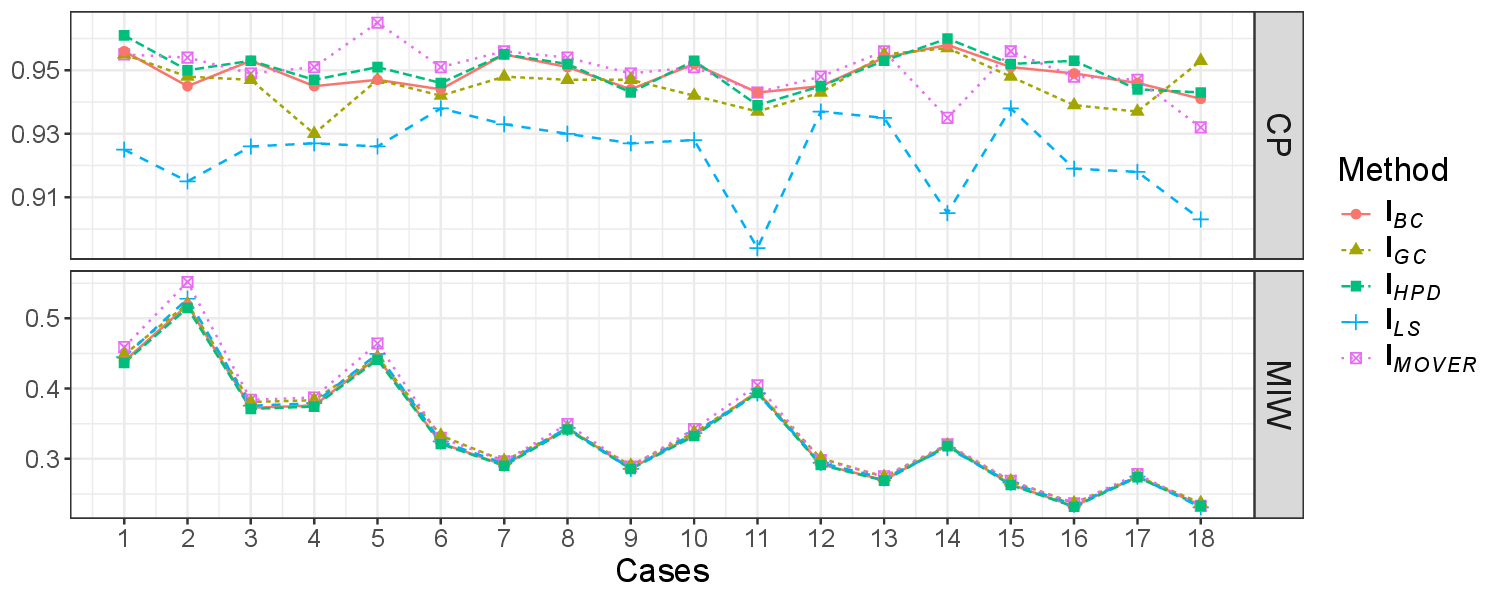}
	\captionof{figure}{CP and MIW of the interval estimators of $\eta$ for $k=2$ and $\alpha=0.05$.}
	\label{fig:1}
	\includegraphics[width=0.95\linewidth, height=0.375\linewidth]{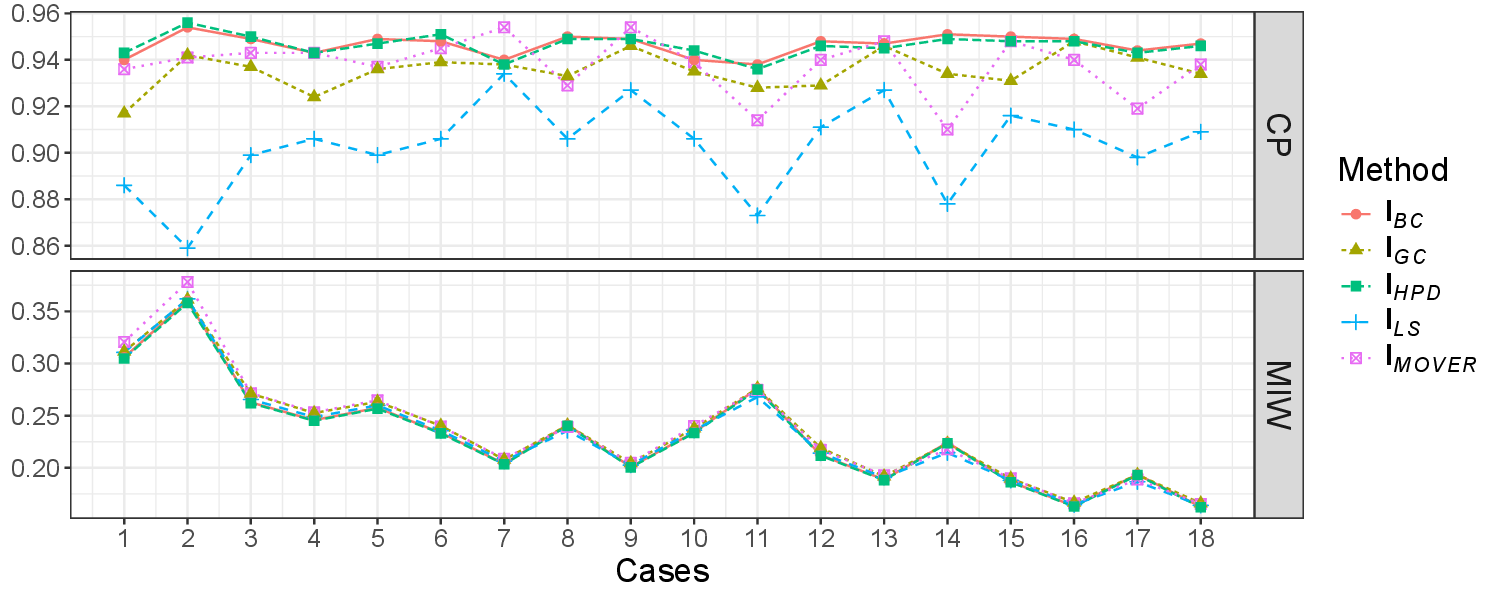}
	\captionof{figure}{CP and MIW of the interval estimators of $\eta$ for $k=4$ and $\alpha=0.05$.}
	\label{fig:2}
\end{figure}
\section{Numerical Monte Carlo Simulation}
In this section, the performances of the various methods proposed in this article are evaluated and compared through a Monte Carlo simulation study. The coverage probability (CP) and mean interval width (MIW) of an interval estimate are chosen as the criteria for evaluating its performance. Several cases have been selected for study for the cases of $k=2$ and $k=4$ populations with the common shape parameter $\eta=0.75$ in all cases. For the case of $k=2$, the two populations have rate parameters $\lambda_1=0.05$ and $\lambda_2 = 0.1$ respectively, while for the latter, the four populations have rates $\lambda_1 = 0.05, \lambda_2 = 0.1, \lambda_3 = 0.15$ and $\lambda_4 = 0.2$ respectively. It is to be noted that $\eta = 0.75 < 1$ corresponds to the case of decreasing density (or equivalently, decreasing failure rate). The simulation study for $\eta \geq 1$ (increasing failure rate) has also been considered, which yields similar results and is thereby not reported in detail. All simulations, computations, and visualizations are done using the R software version 4.6.0. \\
\begin{figure}[b!]
	\centering
	\includegraphics[width=0.95\linewidth, height=0.375\linewidth]{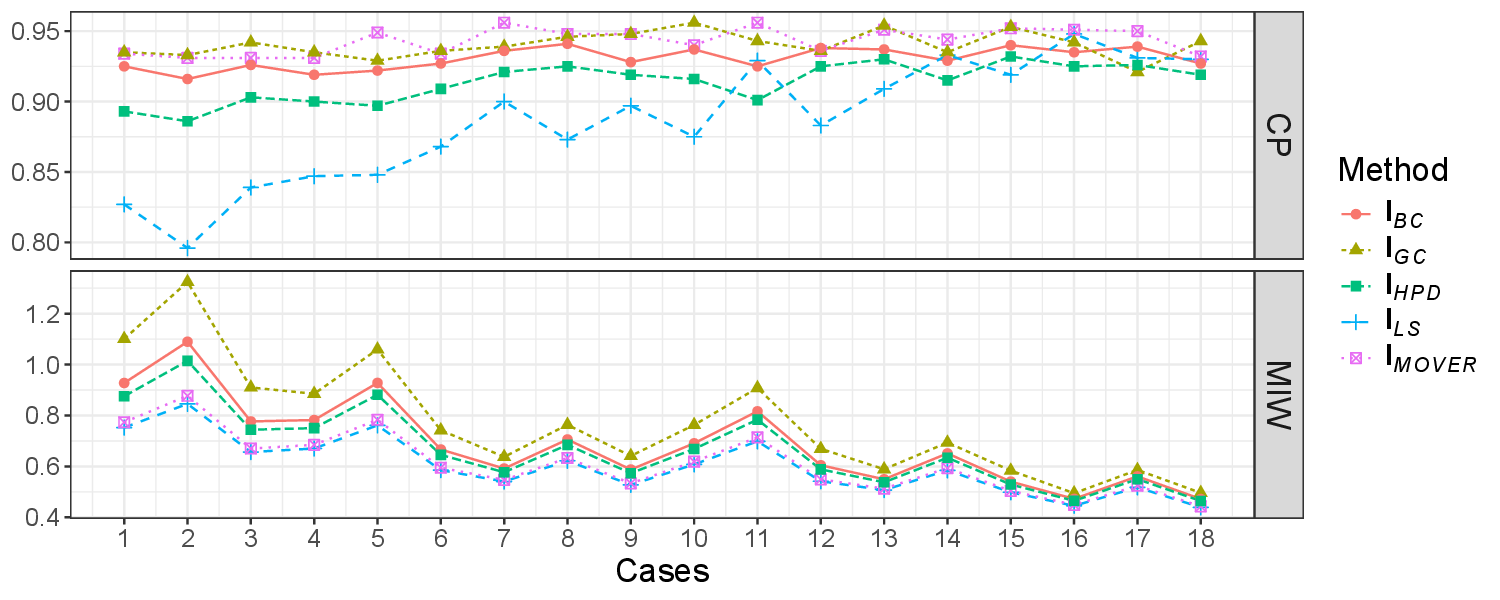}
	\captionof{figure}{CP and MIW of the interval estimators of C$_\eta$ for $k=2$ and $\alpha=0.05$.}
	\label{fig:3}
\end{figure}
\begin{figure}[b!]
	\centering
	\includegraphics[width=0.95\linewidth, height=0.375\linewidth]{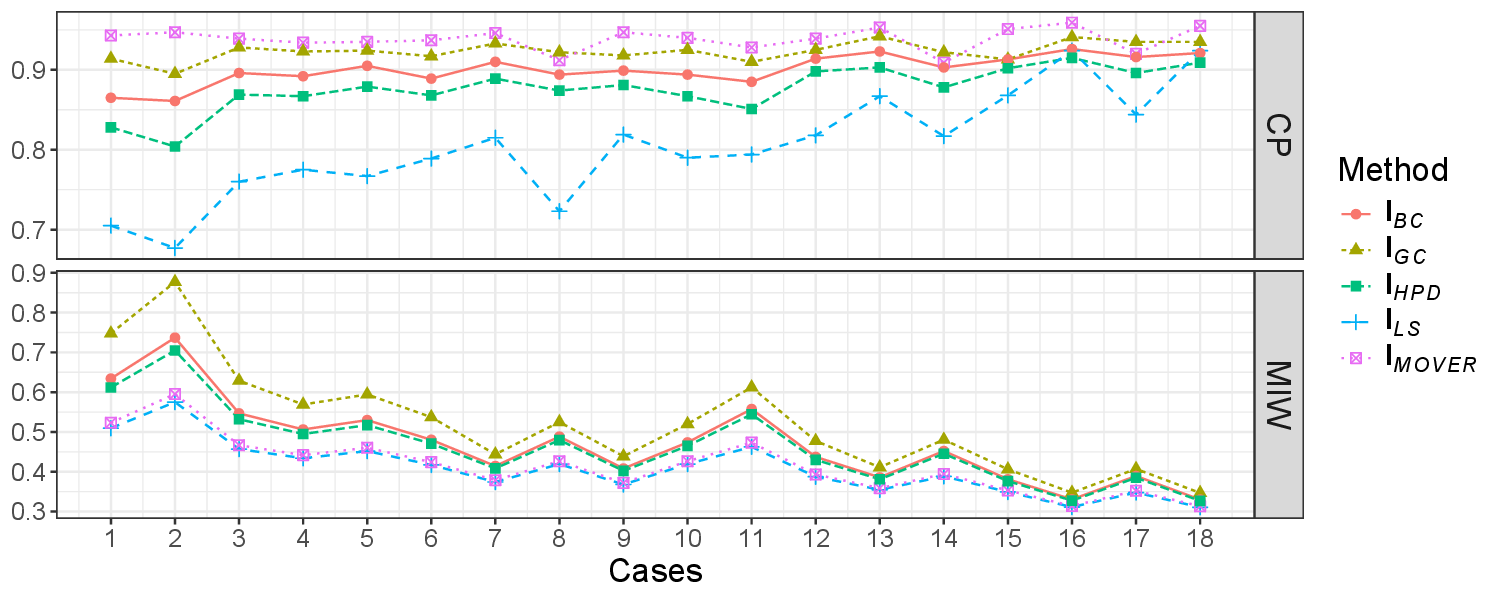}
	\captionof{figure}{CP and MIW of the interval estimators of C$_\eta$ for $k=4$ and $\alpha=0.05$.}
	\label{fig:4}
\end{figure}
A total of 5000 sample replications are generated using the algorithm by \cite{NB1}. Three types of censoring plans are considered. For given sample size $n_j$ and number of failures $m_j$, the first censoring plan involves uniform removal of one surviving unit per failure event until the final observed failure, at which all remaining surviving units are removed. That is $R_{ij}=1~\forall~i=1,2,...,m_j-1$ and $R_{m_jj}=n_j-2m_j+1$. The second plan is the ordinary type-II right censoring plan, where $R_{ij}=0~\forall~i=1,2,...,m_j-1$ and $R_{m_jj}=n_j-m_j$. The third and final plan is the proposed optimal censoring plan, that is the set $\{R_{ij}~:~i=1:m_j;~j=1:k\}$ satisfying equation \eqref{3.3}. A random censoring plan is selected at the initial stage, and the simulated annealing algorithm \cite{DB} is employed. This process starts by iteratively generating neighbouring schemes through redistribution of censored units. At each iteration, a neighbouring scheme is accepted if it improves the objective function \eqref{3.3}. The algorithm is run multiple times using different random initial schemes to account for the stochastic nature of the algorithm. The censoring scheme that provides the smallest objective variance is chosen as the optimal scheme. The initial temperature was selected at 100$^\circ$ with a cooling of 0.995 per iteration until a maximum iteration of 5000 is achieved. 18 cases are organized into three primary categories under the three censoring plans-- uniform, type-II, and optimal censoring plans, respectively, which are listed as follows:
\begin{itemize}
	\item Uniform censoring plan: Cases 1, 4, 7, 10, 13, 16.
	\item Type-II censoring plan: Cases 2, 5, 8, 11, 14, 17.
	\item Optimal censoring plan: Cases 3, 6, 9, 12, 15, 18.
\end{itemize} 
\begin{table}[t!]
	\centering
	\caption{Sampling Plan for various cases}
	\label{tab:my-table-1}
	\begin{tabular}{ccccccc}
		\hline
		Cases       & 1-3      & 4-6                                    & 7-9                                     & 10-12    & 13-15                                   & 16-18     \\ \hline
		Sample size combination & $(30_k)$ & $(30_{\frac{k}{2}}, 50_{\frac{k}{2}})$ & $(30_{\frac{k}{2}}, 100_{\frac{k}{2}})$ & $(50_k)$ & $(50_{\frac{k}{2}}, 100_{\frac{k}{2}})$ & $(100_k)$ \\ \hline
	\end{tabular}
\end{table}
These combinations are further categorized through various selections of sample sizes $n_j=30,50,100$, and upon setting the number of failures to $m_j =  \frac{n_j}{2}$, the structure is given in Table \ref{tab:my-table-1}. Here, $(a_x, b_y)$ represents the tuple $(a,a,...,\text{to}~x~\text{times}, b,b,...,\text{to}~y~\text{times})$. The confidence intervals are obtained at $\alpha=0.05$ or 95\% level of confidence. For the $\textbf{I}_{GC}$, $B = 1000$ pivotal quantities are generated for both the common shape parameter $\eta$ and $\text{C}_\eta$. For the Bayesian intervals, the number of runs for the slice sampling algorithm is set at $B=20000$. The tuning parameters are set at $M=20$ and $w=0.01$ for all cases. The hyperparameters are obtained by equating the prior means to the setting parameters. The results in terms of CP and MIW are plotted in Figures \ref{fig:1}-\ref{fig:4}.\\
\begin{sidewaystable}[!]
	\centering
	\caption{The 95\% confidence and credible intervals for $\eta$ using real data}
	\label{tab:my-table-2}
	\begin{tabular*}{\textheight}{@{\extracolsep\fill}ccccccc}
		\hline
		\multirow{2}{*}{$(m_1, m_2)$}                 & \multirow{2}{*}{Scheme}  & \multirow{2}{*}{$\textbf{I}_{LS}$}              & \multirow{2}{*}{$\textbf{I}_{MOVER}$}            & \multirow{2}{*}{$\textbf{I}_{GC}$}              & \multirow{2}{*}{$\textbf{I}_{BC}$}              & \multirow{2}{*}{$\textbf{I}_{HPD}$}             \\
		&                          &                                   &                                   &                                   &                                   &                                   \\ \hline
		\multirow{6}{*}{(34, 31)} & \multirow{2}{*}{Uniform} & \multirow{2}{*}{(2.9879, 4.5032)} & \multirow{2}{*}{(3.0866, 4.622)}  & \multirow{2}{*}{(3.3173, 4.7490)} & \multirow{2}{*}{(2.9853, 4.3310)} & \multirow{2}{*}{(2.9852, 4.3311)} \\
		&                          &                                   &                                   &                                   &                                   &                                   \\  
		& \multirow{2}{*}{Type-II} & \multirow{2}{*}{(2.8032, 4.8379)} & \multirow{2}{*}{(2.9683, 5.0491)} & \multirow{2}{*}{(3.4277, 5.4043)} & \multirow{2}{*}{(3.1597, 4.9851)} & \multirow{2}{*}{(3.1672, 4.9859)} \\
		&                          &                                   &                                   &                                   &                                   &                                   \\  
		& \multirow{2}{*}{OC}      & \multirow{2}{*}{(2.9479, 4.4528)} & \multirow{2}{*}{(2.8532, 6.6801)} & \multirow{2}{*}{(3.3222, 4.7706)} & \multirow{2}{*}{(3.0816, 4.4359)} & \multirow{2}{*}{(3.0816, 4.4361)} \\
		&                          &                                   &                                   &                                   &                                   &                                   \\ \hline
		\multirow{6}{*}{(14, 13)}                     & \multirow{2}{*}{Uniform} & \multirow{2}{*}{(2.4399, 6.0781)} & \multirow{2}{*}{(2.8532, 6.6801)} & \multirow{2}{*}{(2.7909, 6.4043)} & \multirow{2}{*}{(2.5234, 5.3644)} & \multirow{2}{*}{(2.3917, 5.1991)} \\
		&                          &                                   &                                   &                                   &                                   &                                   \\  
		& \multirow{2}{*}{Type-II} & \multirow{2}{*}{(3.2917, 6.8781)} & \multirow{2}{*}{(3.6883, 7.4203)} & \multirow{2}{*}{(2.9078, 6.5876)} & \multirow{2}{*}{(2.4300, 5.1904)} & \multirow{2}{*}{(2.3873, 5.1439)} \\
		&                          &                                   &                                   &                                   &                                   &                                   \\  
		& \multirow{2}{*}{OC}      & \multirow{2}{*}{(2.8419, 5.0454)} & \multirow{2}{*}{(3.0118, 5.2655)} & \multirow{2}{*}{(3.1290, 5.4347)} & \multirow{2}{*}{(2.4541, 4.2173)} & \multirow{2}{*}{(2.4296, 4.1870)} \\
		&                          &                                   &                                   &                                   &                                   &                                   \\ \hline
	\end{tabular*}
	\vspace{0.5cm}
	\caption{The 95\% confidence and credible intervals for $\text{C}_\eta$ using real data}
	\label{tab:my-table-3}
	\begin{tabular*}{\textheight}{@{\extracolsep\fill}ccccccc}
		\hline
		\multirow{2}{*}{$(m_1, m_2)$}                 & \multirow{2}{*}{Scheme}  & \multirow{2}{*}{$\textbf{I}_{LS}$}              & \multirow{2}{*}{$\textbf{I}_{MOVER}$}            & \multirow{2}{*}{$\textbf{I}_{GC}$}              & \multirow{2}{*}{$\textbf{I}_{BC}$}              & \multirow{2}{*}{$\textbf{I}_{HPD}$}             \\
		&                          &                                   &                                   &                                   &                                   &                                   \\ \hline
		\multirow{6}{*}{(34, 31)} & \multirow{2}{*}{Uniform} & \multirow{2}{*}{(0.2425, 0.3425)} & \multirow{2}{*}{(0.2481, 0.3491)} & \multirow{2}{*}{(0.2395, 0.3359)} & \multirow{2}{*}{(0.2605, 0.3673)} & \multirow{2}{*}{(0.2569, 0.3604)} \\
		&                          &                                   &                                   &                                   &                                   &                                   \\  
		& \multirow{2}{*}{Type-II} & \multirow{2}{*}{(0.2230, 0.3216)} & \multirow{2}{*}{(0.2377, 0.3374)} & \multirow{2}{*}{(0.2027, 0.3098)} & \multirow{2}{*}{(0.2225, 0.3395)} & \multirow{2}{*}{(0.2186, 0.3339)} \\
		&                          &                                   &                                   &                                   &                                   &                                   \\  
		& \multirow{2}{*}{OC}      & \multirow{2}{*}{(0.2354, 0.3234)} & \multirow{2}{*}{(0.2401, 0.3289)} & \multirow{2}{*}{(0.2225, 0.3240)} & \multirow{2}{*}{(0.2362, 0.3381)} & \multirow{2}{*}{(0.2329, 0.3337)} \\
		&                          &                                   &                                   &                                   &                                   &                                   \\ \hline
		\multirow{6}{*}{(14, 13)}                     & \multirow{2}{*}{Uniform} & \multirow{2}{*}{(0.1111, 0.2160)} & \multirow{2}{*}{(0.2275, 0.3357)} & \multirow{2}{*}{(0.1248, 0.2662)} & \multirow{2}{*}{(0.1799, 0.3629)} & \multirow{2}{*}{(0.1710, 0.3478)} \\
		&                          &                                   &                                   &                                   &                                   &                                   \\  
		& \multirow{2}{*}{Type-II} & \multirow{2}{*}{(0.1133, 0.2153)} & \multirow{2}{*}{(0.1839, 0.2887)} & \multirow{2}{*}{(0.1370, 0.3000)} & \multirow{2}{*}{(0.1950, 0.3899)} & \multirow{2}{*}{(0.1902, 0.3799)} \\
		&                          &                                   &                                   &                                   &                                   &                                   \\  
		& \multirow{2}{*}{OC}      & \multirow{2}{*}{(0.2105, 0.3416)} & \multirow{2}{*}{(0.2239, 0.3575)} & \multirow{2}{*}{(0.2106, 0.3565)} & \multirow{2}{*}{(0.2393, 0.4105)} & \multirow{2}{*}{(0.2294, 0.3965)} \\
		&                          &                                   &                                   &                                   &                                   &                                   \\ \hline
	\end{tabular*}
\end{sidewaystable}
It is observed that the simulation results for all $k=2, 4$ show almost the same trend, with $k=2$ exhibiting better coverage but smaller precision than the latter. With an increase in sample sizes across the combinations, the CP values for $\text{C}_\eta$ also tend to increase in almost all cases, especially for the large sample intervals; however, this trend is not observed for $\eta$. Meanwhile, an increase in sample sizes across the combinations reduces the MIW values in almost all cases for both $\eta$ and $\text{C}_\eta$. In terms of the censoring plans, it is observed for $\eta$ that the optimal censoring plan yields the smallest MIW in all cases, and also the highest CP values for smaller sample size combinations. These trends are, in fact, even more favourable for $\text{C}_\eta$, where the optimal censoring plan dominates in both coverage and precision. \\
\indent Now the comparisons among the performance of the proposed methods are made and discussed. In terms of CP values for $\eta$, the $\textbf{I}_{MOVER}$ intervals are in contention with the $\textbf{I}_{BC}$ and $\textbf{I}_{HPD}$ for exhibiting the highest coverage. It is also observed that $\textbf{I}_{LS}$ performs the weakest in all cases. In terms of the MIW values for $\eta$, $\textbf{I}_{HPD}$ has the greatest precision among the proposed interval estimators. Meanwhile, it is observed for $\text{C}_\eta$ that the $\textbf{I}_{MOVER}$ and $\textbf{I}_{GC}$ have the highest CP values in almost all cases. Although the interval estimator $\textbf{I}_{LS}$ appears to have the smallest MIW in all cases, its precision is, however, countered by its weak coverage. Overall, $\textbf{I}_{MOVER}$ has the most satisfactory performance for $\text{C}_\eta$, while $\textbf{I}_{HPD}$ does so for $\eta$, due to their high CP and small MIW values.
\section{Carbon Fiber Strength Data Example}
\begin{figure*}[t!]
	\centering
	\begin{subfigure}[t]{0.5\textwidth}
		\centering
		\includegraphics[width=0.9\linewidth]{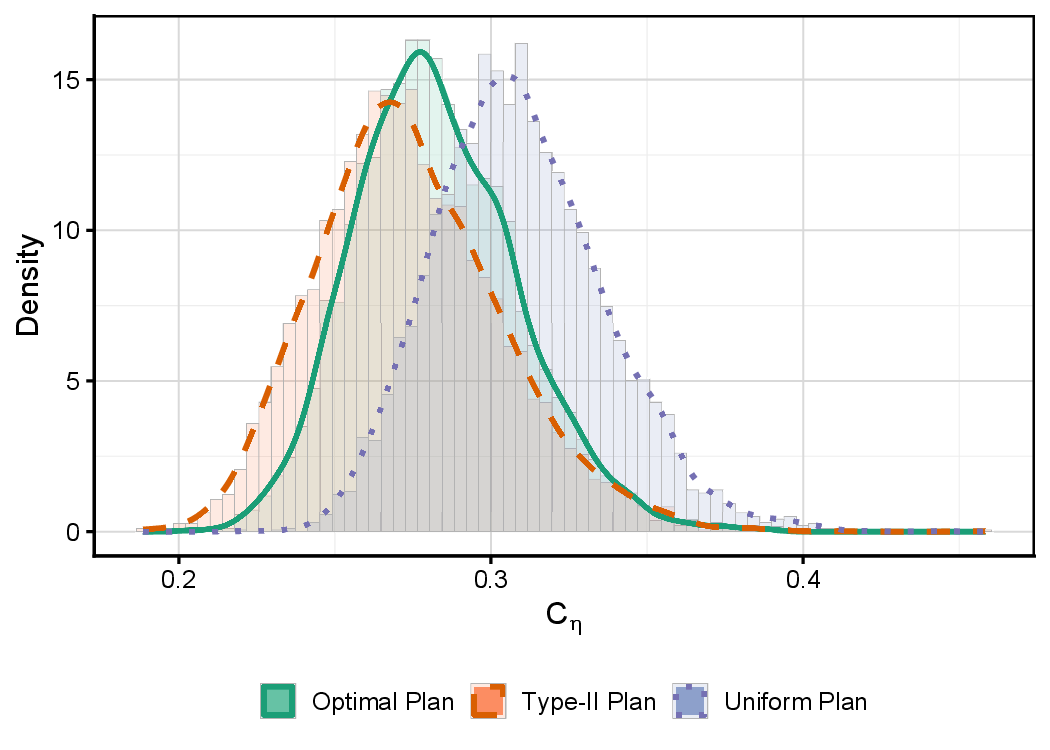}
		\caption{Posterior distributions of $\text{C}_\eta$}
		\label{fig:8a}
	\end{subfigure}%
	\begin{subfigure}[t]{0.5\textwidth}
		\centering
		\includegraphics[width=0.9\linewidth]{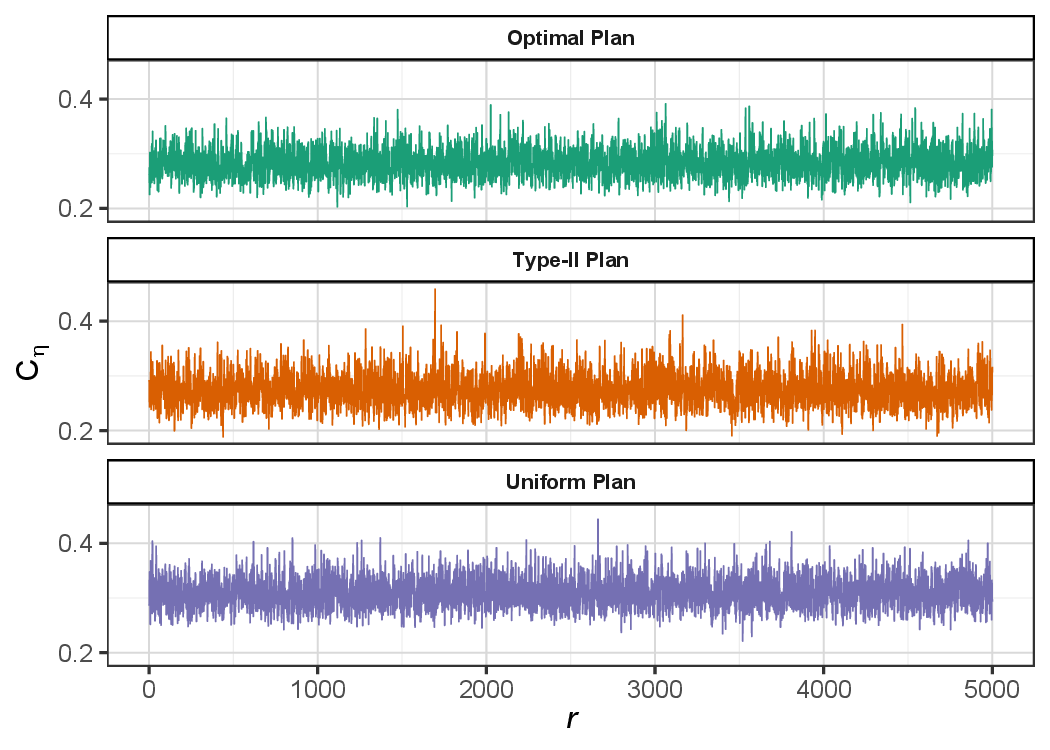}
		\caption{Trace plots of $\text{C}_\eta$}
		\label{fig:8b}
	\end{subfigure}
	\caption{MCMC diagnostic plots of $\text{C}_\eta$ under 50\% censoring using real data}
	\label{fig:8}
	\centering
	\begin{subfigure}[t]{0.5\textwidth}
		\centering
		\includegraphics[width=0.9\linewidth]{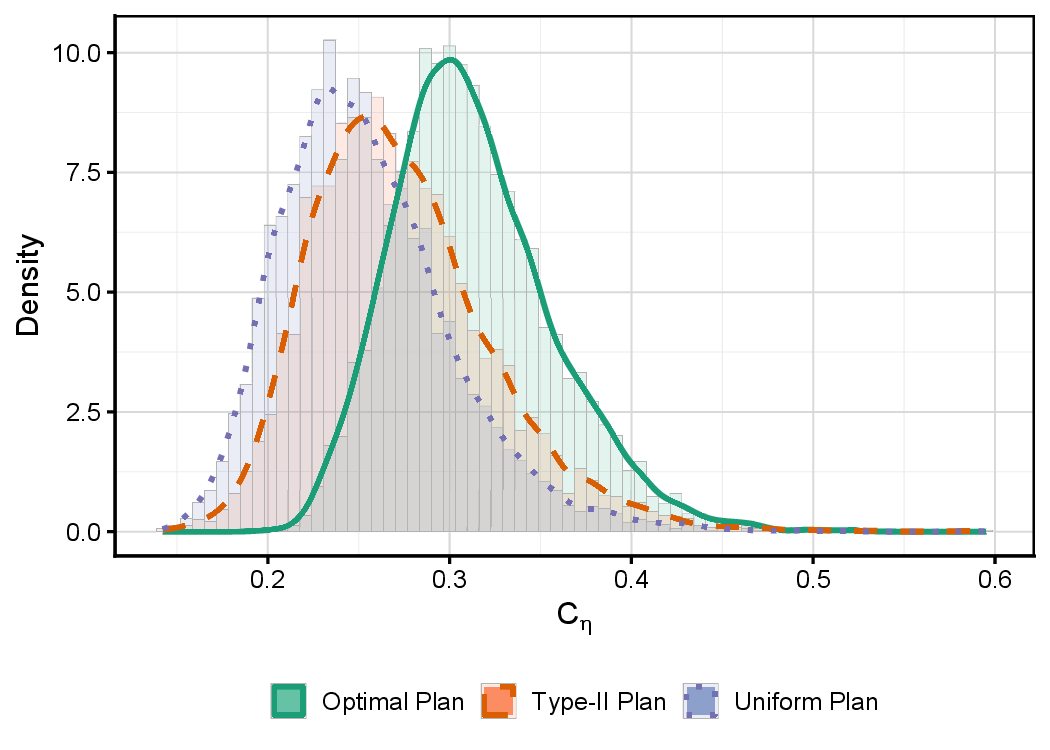}
		\caption{Posterior distributions of $\text{C}_\eta$}
		\label{fig:9a}
	\end{subfigure}%
	\begin{subfigure}[t]{0.5\textwidth}
		\centering
		\includegraphics[width=0.9\linewidth]{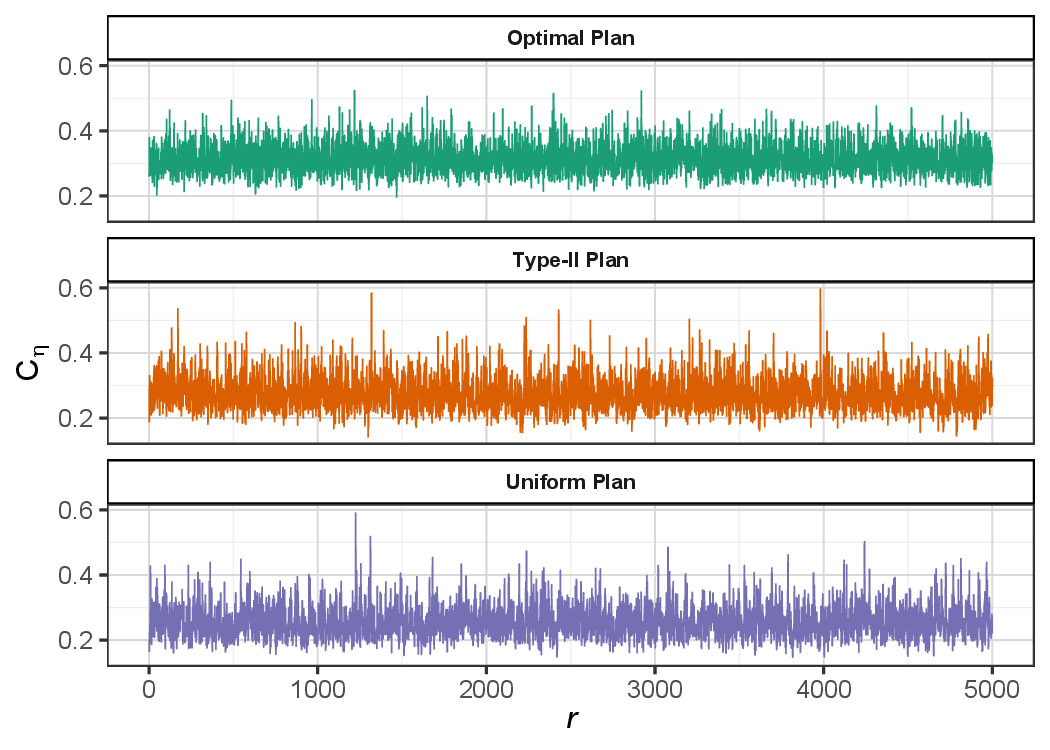}
		\caption{Trace plots of $\text{C}_\eta$}
		\label{fig:9b}
	\end{subfigure}
	\caption{MCMC diagnostic plots of $\text{C}_\eta$ under 80\% censoring using real data}
	\label{fig:9}
\end{figure*}

In this section, an illustrative example of the proposed methodologies using real data given in \cite{MGB} is discussed. This data relates to the strength of two types of single carbon fibers and impregnated 1000 carbon fiber tows, measured in gigapascal. These fibers were kept under tension for testing at 20 mm and 10 mm gauge lengths. It has already been shown by \cite{DK}	that the Weibull distributions with equal shape parameters fit to both of these data sets upon subtracting 0.75 from them. See further details for testing and inference in \cite{SM} and \cite{FS}. \\
\begin{figure*}[t!]
	\centering
	\begin{subfigure}[t]{0.45\textwidth}
		\centering
		\includegraphics[width=0.8\linewidth]{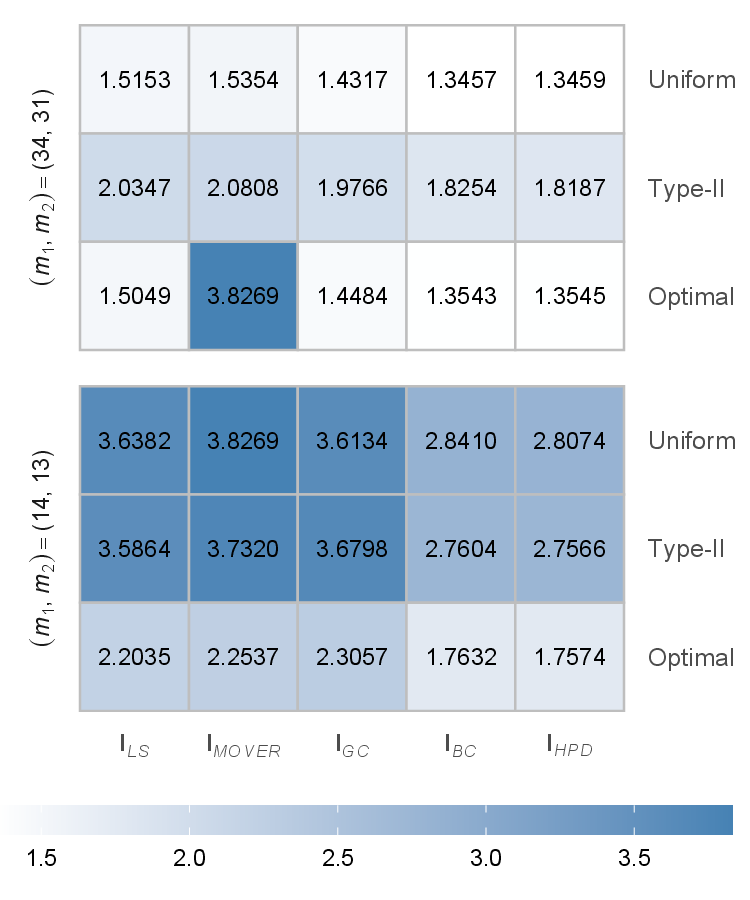}
		\caption{$\eta$}
	\end{subfigure}%
	\begin{subfigure}[t]{0.45\textwidth}
		\centering
		\includegraphics[width=0.8\linewidth]{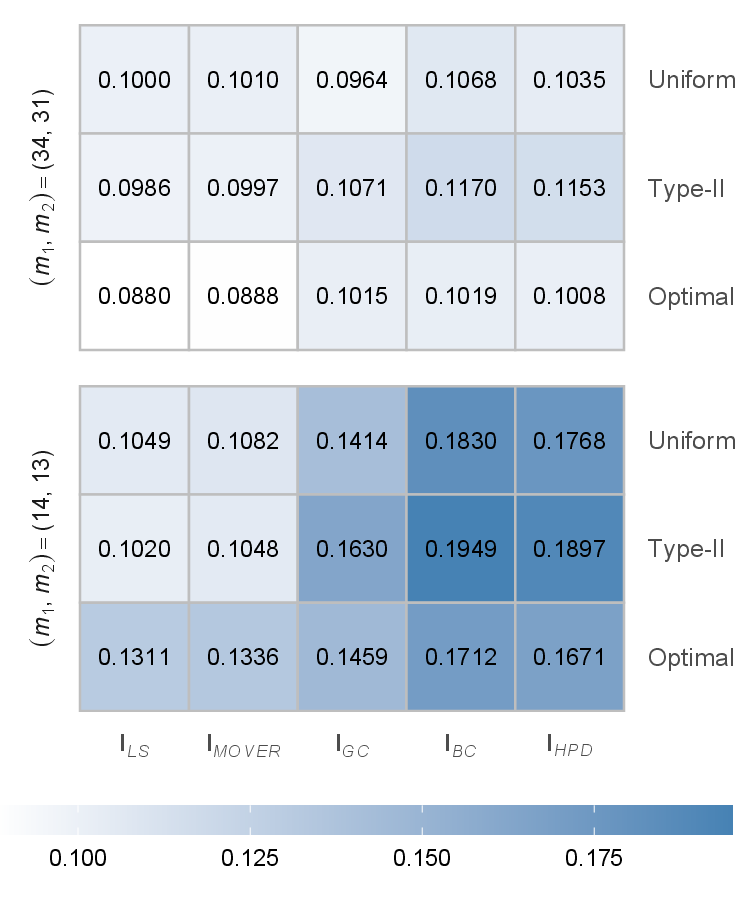}
		\caption{$\text{C}_\eta$}
	\end{subfigure}
	\caption{Widths of the interval estimates of $\eta$ and $\text{C}_\eta$ for real dataset.}
	\label{fig:5}
\end{figure*}
\begin{figure*}[b!]
	\centering
	\begin{subfigure}[b]{0.5\textwidth}
		\centering
		\includegraphics[width=0.9\linewidth]{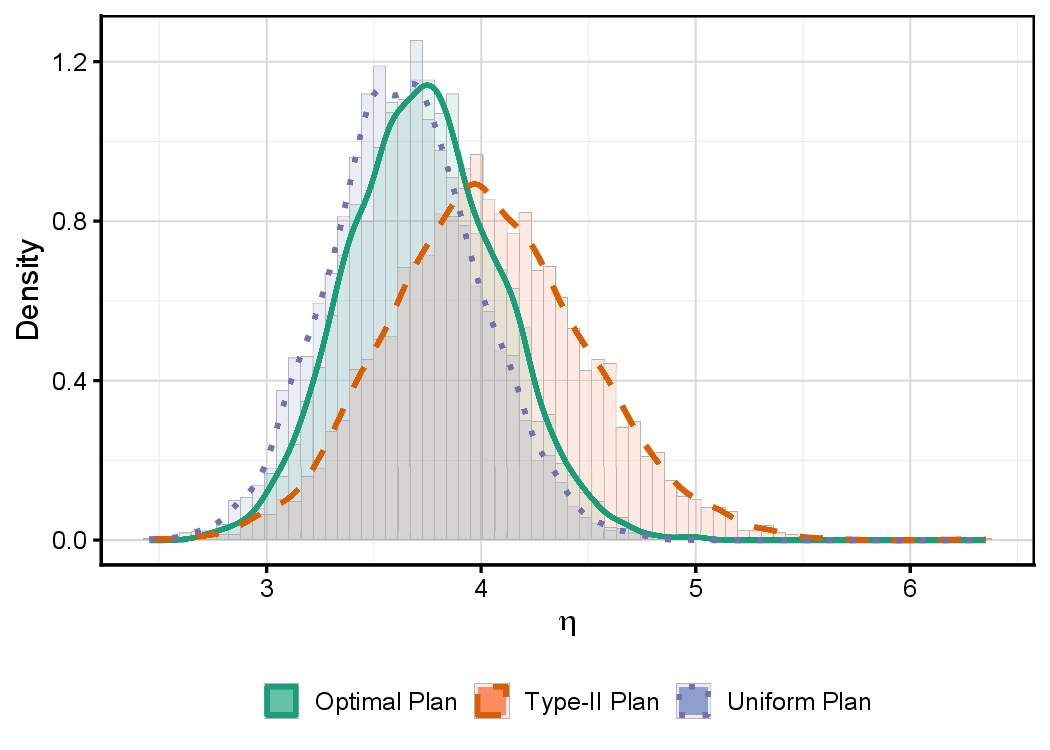}
		\caption{Posterior distributions of $\eta$}
	\end{subfigure}%
	\begin{subfigure}[b]{0.5\textwidth}
		\centering
		\includegraphics[width=0.9\linewidth]{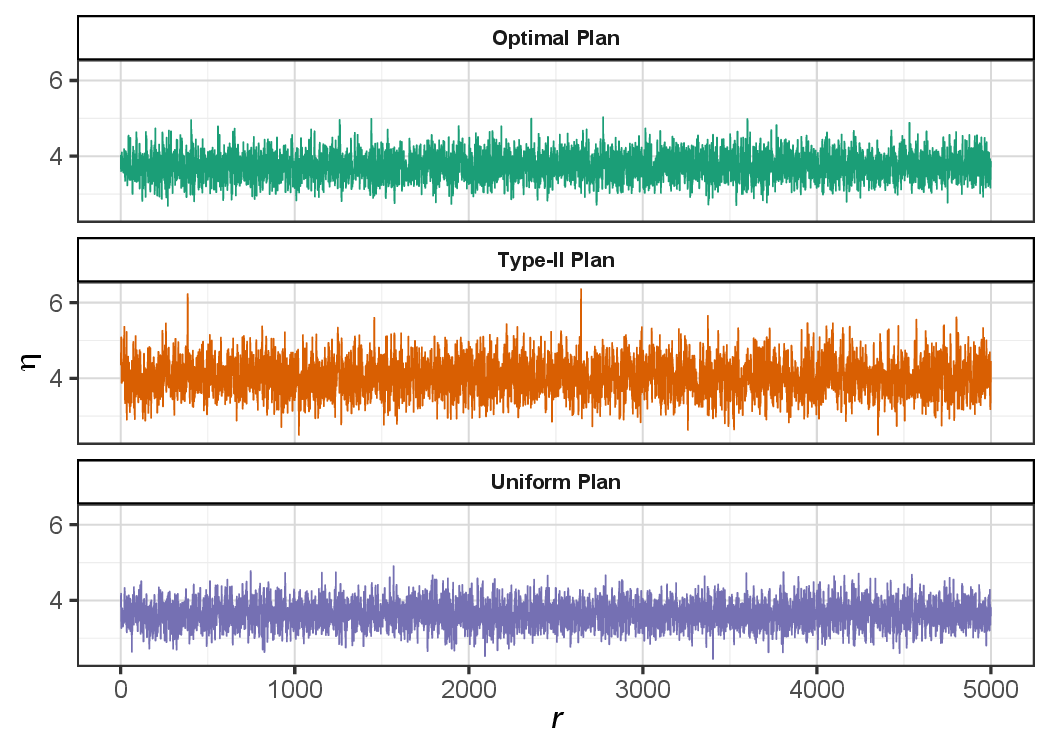}
		\caption{Trace plots of $\eta$}
	\end{subfigure}
	\caption{MCMC diagnostic plots of $\eta$ under 50\% censoring using real data}
	\label{fig:6}
	\centering
	\begin{subfigure}[t]{0.5\textwidth}
		\centering
		\includegraphics[width=0.9\linewidth]{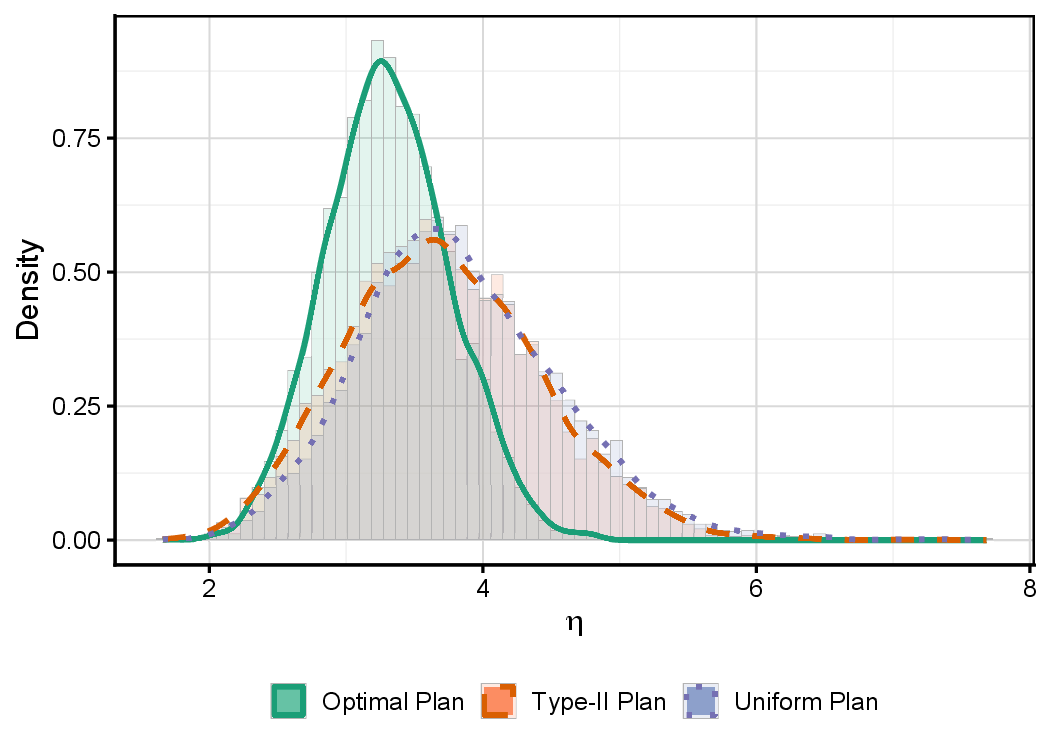}
		\caption{Posterior distributions of $\eta$}
	\end{subfigure}%
	\begin{subfigure}[t]{0.5\textwidth}
		\centering
		\includegraphics[width=0.9\linewidth]{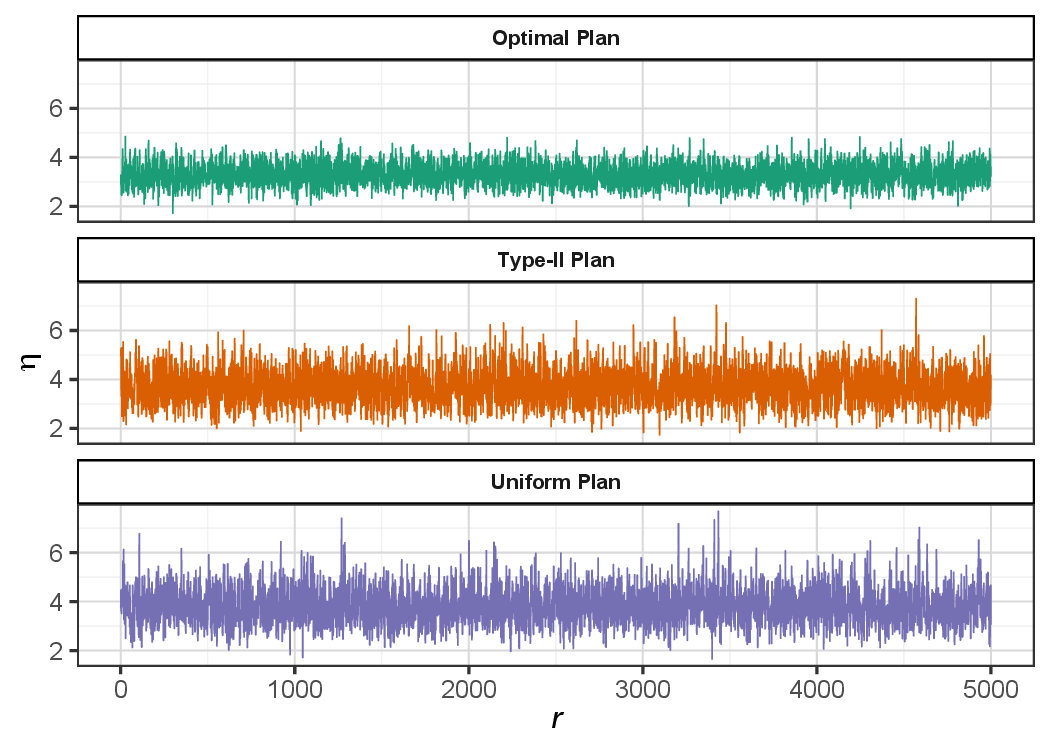}
		\caption{Trace plots of $\eta$}
	\end{subfigure}
	\caption{MCMC diagnostic plots of $\eta$ under 80\% censoring using real data}
	\label{fig:7}
\end{figure*}
The 95\% confidence and credible intervals have been calculated under the consideration of the same three censoring schemes - Uniform, type-II and optimal censoring scheme, as discussed in the numerical simulation Section 4, for these $k=2$ samples. To illustrate the behavior of the interval estimates with respect to the number of failures, two scenarios are considered at 50\% and 80\% censoring. The first case corresponds to $(m_1, m_2) = (\lfloor\frac{n_1}{2}\rfloor, \lfloor\frac{n_2}{2}\rfloor) = (34, 31)$ while the latter corresponds to $(m_1, m_2) = (14, 13)$.\\
All the methodologies are applied for these cases. For the $\textbf{I}_{GC}$, the number of generalized pivotal items are set at $B=1000$, while for the Bayesian methods, the overall MCMC iteration was run up to $\mathscr{R}=50000$. The MLEs have been used as initial approximations, and therefore no burn-in of samples was done. Furthermore, to counter for the high auto-correlation between samples, a thinning process of selecting 1 in every 50 samples is considered, reducing the overall size to 5000. Due to unavailability of prior information, the hyperparameters are treated as zero resulting in the non-informative Jeffreys' prior. The slice sampling was tuned using the standard practice of setting $M=20$, with a stepping out increment of $w=0.1$. The results are reported in the Tables \ref{tab:my-table-2} and \ref{tab:my-table-3}. To compare the widths, the results are displayed in Figure \ref{fig:5}. The MCMC diagnostic plots are also given in Figures \ref{fig:6}--\ref{fig:9}.\\
\indent As evident from Figure \ref{fig:5}, the optimal censoring scheme generally yields the smallest interval widths, although a few exceptions are seen in the instances of $(m_1,m_2)=(34, 31)$ for $\textbf{I}_{MOVER}$. This exception is also observed in the increase of the number of failures. Meanwhile in every other case, irrespective of the methods, an increase in the values of $(m_1, m_2)$ significantly reduce the widths of the estimates, as more information is brought in. Among the methods, the $\textbf{I}_{HPD}$ have the smallest interval widths in all cases for $\eta$. For $\text{C}_\eta$, the $\textbf{I}_{LS}$ are the narrowest, except for the case of uniform censoring with $(m_1, m_2) = (34, 31)$, where the $\textbf{I}_{GC}$ is the smallest in width.
\section{Conclusion}\label{sec5}
In this article, the problem of estimating the common shape parameter and coefficient of variation of several ($k$) populations following the Weibull distribution with unequal rate parameters under progressive type-II censoring is considered. Since this choice of censoring scheme generalizes the complete sampling plan (for $n_j = m_j$ and $R_{ij} = 0$ for all $i=1,2,...,m_j$ and $j=1,2,...,k$), this work presents a unified and generalized framework that marks some of the results by \cite{MLo1,MLB} as special cases. The proposed methodology includes confidence interval methods based on large-sample theory, variance-estimate recovery, and generalized pivotal quantities. The Bayesian credible intervals such as the equal-tailed and highest posterior density intervals are also constructed using a hybrid Slice-within-Gibbs sampling algorithm. Furthermore, to enhance robustness of the methods, an optimal censoring design is devised and compared with other popular progressive type-II censored plans. It was found that from a comprehensive simulation study, the confidence intervals based on the method of variance-estimate recovery and the Bayesian framework exhibit the highest coverage, with the highest posterior density interval yielding the most satisfactory precision for the common shape parameter. For the common coefficient of variation, the confidence interval based on the method of variance estimates recovery has the highest coverage probability. A real carbon fiber strength dataset was also used to illustrate the applicability of the proposed methodology. The problem considered in this article can be further extended to several dependent populations and under adaptive censoring schemes. The generalization can also be considered for frailty parameters or proportional hazards families. \\\\
\noindent {\bf{\large Declarations}}\\
\noindent {\bf Availability of data and materials}:	The data used in the present study are obtained from a previously published article, which is appropriately cited in the manuscript. The data are available from the corresponding published source.\\
\noindent {\bf Competing interest}: The authors declare that they have no competing interests.\\
\noindent {\bf Funding}: No funding was received for conducting this research.\\
\noindent {\bf Authors' contributions}: The first author contributed to the conceptualization of the study, development of the methodology, statistical analysis, performed the computational and simulation studies, prepared the first draft of the manuscript, interpretation of the results, and critical revision of the manuscript. 
The second author contributed to the supervision, conceptualization and development of the methodology, performed the computational and simulation studies, prepared the first draft of the manuscript, and contributed to its subsequent revision. Both authors read and approved the final manuscript.\\
\noindent {\bf Acknowledgement}: Not applicable.
	\bibliographystyle{unsrt}
	\bibliography{references}
\end{document}